\documentclass[fleqn,usenatbib,usedcolumn]{mnras}
\usepackage[british]{babel}
\usepackage{newtxtext}
\usepackage[slantedGreek]{newtxmath}
\usepackage[T1]{fontenc}
\usepackage{graphicx, mathrsfs}

\newcommand\lsim{\mathrel{\rlap{\lower4pt\hbox{\hskip1pt$\sim$}}
        \raise1pt\hbox{$<$}}}
\newcommand\gsim{\mathrel{\rlap{\lower4pt\hbox{\hskip1pt$\sim$}}
        \raise1pt\hbox{$>$}}} 

\usepackage{times}
\usepackage{graphicx}
\usepackage{verbatim}
\usepackage{tikz}
\usepackage{color}

\usepackage[T1]{fontenc}
\usepackage{aecompl}
\usepackage{calligra}
\DeclareMathAlphabet{\mathcalligra}{T1}{calligra}{m}{n}
\DeclareFontShape{T1}{calligra}{m}{n}{<->s*[2.2]callig15}{}

\usepackage{hyperref}
\hypersetup{
    pdfnewwindow=true,      % links in new window
    colorlinks=true,       % false: boxed links; true: colored links
    linkcolor=blue,          % color of internal links
    citecolor=cyan,        % color of links to bibliography
    filecolor=blue,      % color of file links
    urlcolor=blue           % color of external links
}

\def\prd{PRD}
\def\apj{ApJ}                 % Astrophysical Journal
\def\apjl{ApJL}               % Astrophysical Journal, Letters
\def\apjs{ApJS}               % Astrophysical Journal, Supplement
\def\mnras{MNRAS}             % Monthly Notices of the RAS
\def\aap{A\&A}                % Astronomy and Astrophysics
\def\nat{Nature}              % Nature
\def\araa{ARA\&A}             % Annual Review of Astronomy and Astrophysics
\def\eg{\textit{e.g.}}

\def\orb{\rm{orb}}

\def\max{\rm{max}}

\def\Msun{{M_{\odot}}}

\def\Mc{\mathcal{M}}

\def\bbeta{\boldsymbol{\beta}}
\def\btheta{\boldsymbol{\theta}}
\def\balpha{\boldsymbol{\alpha}}

\def\esc{\rm{esc}}

\def\fint{f_{\orb,0}}

\begin{document}

\title[Eccentric BBHs in the LISA Band]{Black Hole Mergers From Globular Clusters Observable by LISA II: Resolved Eccentric Sources and the Gravitational Wave Background}
\author[D. J. D'Orazio, Johan Samsing]{Daniel J. D'Orazio$^1$\thanks{daniel.dorazio@cfa.harvard.edu; jsamsing@princeton.edu},
  Johan Samsing$^2$ \\
     $^1$Department of Astronomy, Harvard University, 60 Garden Street Cambridge, MA 01238, USA\\
     $^2$Department of Astrophysical Sciences, Princeton University, Peyton Hall, 4 Ivy Lane, Princeton, NJ 08544, USA}

\maketitle
\begin{abstract}

In paper I of this series we showed that a large percentage of the binary
black hole (BBH) mergers that form through dynamical interactions in globular clusters
will have significant eccentricity in the $\sim 10^{-3}-10^{-1}$~Hz LISA band.
In this work we quantify the evolution of these highly eccentric binaries
through the LISA and LIGO bands, and compute the stochastic gravitational wave
background from the merging, eccentric population. We find that the population
of BBHs that merge in-between three-body encounters inside their cluster
($\sim 50\%$ of all cluster-formed BBH mergers) will have measurable
eccentricity for their entire lifetime in the LISA band. The population of
BBHs that merge during three-body encounters ($\sim 5\%$ of all cluster-formed
BBH mergers), will be detectable by LIGO with eccentricities of $e\sim0.1$.
The gravitational wave background from dynamically assembled BBHs encodes a
characteristic bump due to the high initial eccentricities of these systems.
The location and amplitude of this bump depends on globular cluster
properties.

\end{abstract}

\begin{keywords} % commenting results in missing "thanks notes"!
gravitational waves, stars: black holes, galaxies: star clusters: general
\end{keywords}

\maketitle

\section{Introduction}

The Laser Interferometer Gravitational Wave Observatory (LIGO) is beginning to
reveal the population of stellar-mass black holes (BHs) that come together and
merge within a Hubble time \citep{GW150914, GW151226, LIGO_BBHO1:2016,
GW170104, GW170608, GW170814}. To varying degrees of accuracy, LIGO can
measure BH masses, spins, mass ratios, and eccentricities
\citep[\eg][]{2016ApJ...818L..22A, 2016PhRvL.116x1102A}. The distribution of
these parameters for binary black holes (BBHs) at merger will help to quantify
the population of BBHs in the universe, and begin to teach us about the origin
of these systems \citep[\textit{e.g.},][]{OlearyKocsis:2009, KocsisLevin:2012,
2014ApJ...784...71S, BreivikRodLISA:2016, 2016ApJ...832L...2R,
2017Natur.548..426F, 2017ApJ...846...82Z, 2017ApJ...840L..14S,
2017arXiv170901660S, 2017arXiv171107452S,Belczynski+2017, 2018ApJ...853..140S,
2018ApJ...854L...9F}.

LIGO, however, is sensitive only to the final seconds of a BBH inspiral and
merger, at gravitational wave (GW) frequencies of 10's of Hz to kHz. The final
inspiral and merger generates the highest amplitude GWs, and is arguably the
most interesting portion of a BBH merger from a gravitational perspective,
encoding tests of gravity in the strong field \citep{2014ARA&A..52..661L,
TestofGR:LIGO:2016, 2016PhRvD..94h4002Y}. Astrophysically, however, it is only
a small snapshot of the BBH lifetime. While precessional dynamics, as
well as combinations of mass, redshift, and effective spin measurements of
BBHs measured in the LIGO band can encode aspects of formation mechanisms
\citep[see, \textit{e.g.},][]{Gerosa+2013, Kesden+2015, Gerosa+2015,
2016ApJ...832L...2R, GerosaBerti:2017, RodPND+2017, Fishbach+2017}, many of the
astrophysical imprints on formation are washed away by the dissipative effects
of the gravitational radiation reaction that brings the binary to coalescence.

A fuller understanding of the processes that govern the formation and inspiral
of BBH systems merging in the LIGO band requires a census of the BBHs earlier
in their lifetimes, when imprints of formation could be strong. One such
imprint, which is rapidly erased by GW emission later in the BBH evolution, is
binary eccentricity \citep[\textit{e.g.},][]{Gondan+2018}. BBHs formed
via dynamical assembly are expected to have non-zero eccentricity
\citep{Benacquista:2002, OlearyKocsis:2009, KocsisLevin:2012,
2014ApJ...784...71S, 2016PhRvD..94h4013C, 2017ApJ...840L..14S,
2017arXiv170901660S, 2017arXiv171107452S, RodPND+2017, Gondan+2017,
2018ApJ...853..140S, 2018MNRAS.476.1548S, 2018ApJ...855..124S,
2018ApJ...853...93R, 2018arXiv180205718R} in contrast with BBHs formed near
zero eccentricity in the field, with the exception of field BBHs that are
driven to merger through secular evolution
\citep[\textit{e.g.},][]{2017ApJ...836...39S, 2018ApJ...853...93R,
2018arXiv180205718R}. Higher eccentricity causes a binary with otherwise the
same orbital frequency to merge in a shorter time \citep{Peters64}. Higher
eccentricity also causes a binary to emit GWs at a spread of GW frequencies
that are peaked at a higher harmonic of its orbital frequency than for a
binary on a circular orbit, which emits GWs only at the second harmonic of its
orbital frequency \citep[\textit{e.g.},][]{PetersMatthews:1963}.

Interestingly, a subset of the BBH mergers detectable by LIGO will also be
detectable earlier in their evolution by the Laser Interferometer Space
Antennae \citep[LISA;][]{LISA:2017}, at lower GW frequencies, $\sim10^{-2}$~Hz
\citep[see][and references therein]{Fregeau+2006, PauSanta:2010,
SesanaLISALIGO:2016, Seto:2016}. Hence, while LIGO will continue to detect
BBHs at merger, LISA holds the exciting prospect of detecting the same BBHs
closer to formation, providing key information for distinguishing BBH
formation channels \citep{Nishizawa+2016, BreivikRodLISA:2016, Nishizawa+2017,
SamsingDOrazioLISAI:2018}. If the proposed DECIGO \citep{DECIGO} or Tian Qin
\citep{TianQin} detectors succeed in filling in the GW frequency gap between
LISA and LIGO, then for some systems, the majority of the BBH evolution can be
tracked from near formation to merger
\citep[\textit{e.g.},][]{SamsingDOrazio:2018a}.

In \cite{SamsingDOrazioLISAI:2018} (hereafter Paper I) we discussed the
mechanisms by which three different channels for the formation of BBHs in
globular clusters (GCs) lead to three distinct modes in the distribution of
binary eccentricities, and hence distinct GW peak frequencies at formation for
each channel. The three types of BBH mergers resulting from dynamical
formation in GCs are
\begin{itemize}
\item The \textit{ejected mergers}: occurring for BBHs that are hardened over a series of binary-single
interactions until they are sufficiently tightly bound that the next binary-single interaction results in ejection of the BBH from
the GC \citep[\textit{e.g.},][]{2016PhRvD..93h4029R, 2017MNRAS.469.4665P, 2017arXiv171107452S}. If the ejected
BBH has high enough eccentricity at ejection it
will merge in a Hubble time \citep{PZMc:2000, 2016PhRvD..93h4029R, 2017arXiv171107452S}. In the following
we loosely refer to an ejected merger as an ejected BBH that merges within a Hubble time and contributes to the present day observable BBH mergers. 
\item The \textit{2-body in-cluster} (hereafter `2-body') mergers: occurring when a BBH has high enough eccentricity to merge in-between binary-single interactions, before being ejected from the GC \citep{2017arXiv171107452S, RodPND+2017}.
\item The \textit{3-body in-cluster} (hereafter `3-body') mergers: occurring when a BBH is formed in a binary-single interaction with such high eccentricity that a `GW capture' merges the binary in the GC, before the third body can disrupt the
orbit \citep{2006ApJ...640..156G, 2014ApJ...784...71S}.
\end{itemize}

In Paper I, we showed that the ejected mergers form with eccentricities above
$\sim 0.85$ and orbital frequencies of $\sim10^{-7}$~Hz. High eccentricity
causes the GW emission to be emitted primarily at higher harmonics of the
orbital frequency than for a circular orbit, at $\sim10^{-4}-10^{-5}$~Hz, just
below the minimum frequencies that can be probed by the LISA band. On the
other extreme, the 3-body mergers form with similar orbital frequencies, but
such high eccentricity $e\sim0.9999$ that their initial GW frequencies peak
high of the LISA band, at $1-10$~Hz \citep{ChenPau:2017,
SamsingDOrazio:2018a}. The 2-body mergers form with eccentricity intermediate
to the ejected and 3-body mergers, causing their peak GW frequency at
formation to fall directly in the LISA band (Paper I).

In this work, we quantify the evolution of orbital eccentricity and GW
emission of these three BBH populations. In \S \ref{sec:Orbital Evolution and
GW Emission} We compute the eccentricity evolution, characteristic GW strain,
and corresponding signal-to-noise ratios of representative BBHs from each of
the three eccentric populations in the LISA and LIGO bands. We show that LISA
will be able to measure the eccentricity of the 2-body BBH population,
distinguishing it from the ejected merger population that has a lower average
eccentricity, marginally measurable by LISA, and from the 3-body population,
which LISA will not detect \cite{SamsingDOrazio:2018a}. We also show that the
3-body population will have a measurable eccentricity in LIGO, meaning that
LISA and LIGO observations should be able to disentangle the three dynamical
formation channels discussed here. Beyond individually resolvable sources, we
present in \S \ref{S:GWB} a calculation of the stochastic
gravitational wave background from these dynamically formed BBH populations,
including a full treatment of binary eccentricity. Finally, in \S
\ref{S:GCdep}, we explore the dependence of our findings on GC properties,
specifically the escape velocity from the cluster.

\section{Orbital Evolution and GW Emission from Individually Resolved BBHs}
\label{sec:Orbital Evolution and GW Emission}

In this section we quantify the orbital evolution of the dynamically assembled
BBH populations and determine their detectability as individually resolved
sources in LISA and LIGO. The calculations presented here take as input the
initial masses, orbital frequencies and eccentricity distributions of BBHs
formed dynamically in GCs. These distributions are derived in Paper I of this
series with the following assumptions. We assume that all BHs have a mass of
$30\Msun$, so that the BBHs have total masses of $60 \Msun$. This is a
simplifying assumption motivated by the propensity for dynamically formed
systems to form near equal mass, high mass binaries
\citep{2016PhRvD..93h4029R}, and by half of the LIGO BBHs having a similar
make-up \citep[\textit{e.g.},][]{2017ApJ...846...82Z}. Unless otherwise stated, we
choose for the fiducial GC properties, an escape velocity of $v_{\rm esc} =
50$~km s$^{-1}$ and a core single-BH density of $n_{\rm s} = 10^{5}$
pc$^{-3}$. Further details are available in Paper I.

\subsection{Eccentricity evolution}
\label{sec:Eccentricity evolution}

\begin{figure}
\begin{center}
\includegraphics[scale=0.23]{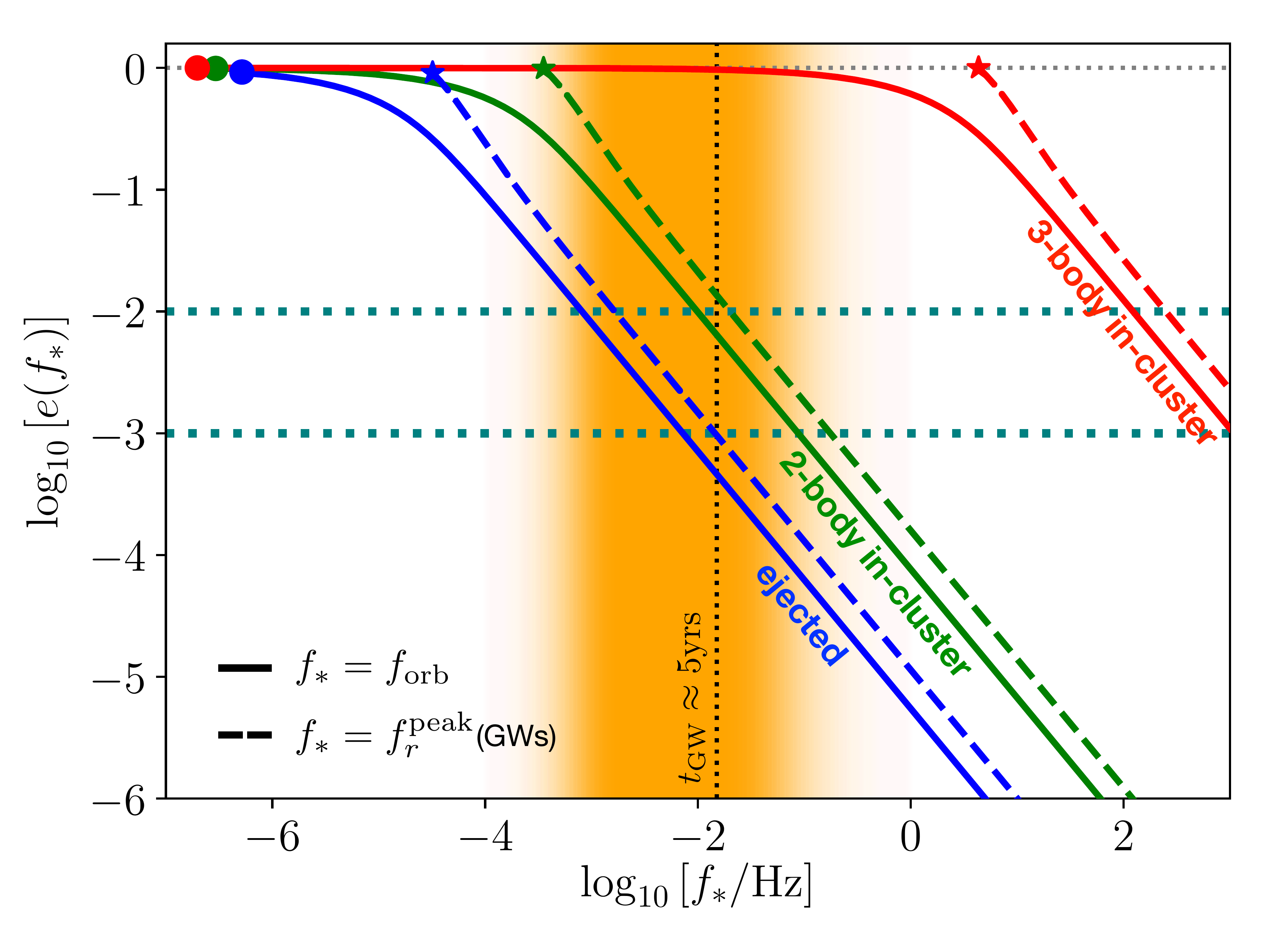}
\end{center}
\vspace{-15pt}
\caption{
The eccentricity evolution of the three BBH populations discussed here and in
Paper I, assuming a GC with an escape velocity of $50$~km~s$^{-1}$. We track
representative BBHs from each population by using the average initial
eccentricity and orbital frequency for each population. The solid lines
represent the eccentricity as a function of binary {\it orbital frequency},
while the dashed lines plot the eccentricity vs. the (peak) {\it gravitational
wave frequency} at which the most power is emitted. The circles denote the
average eccentricity and orbital frequency at which the binaries form and the
stars indicate the corresponding frequency of peak GW emission at formation.
The orange shading represents the relative LISA sensitivity (darkest orange is
the highest sensitivity) and the dashed, horizontal lines are the values above
which LISA can measure the eccentricity of a resolved source in two different
cases (see text). The vertical dotted line denotes the frequency above which
BBHs at $z=0.1$ will be observed to merge within the timescale of the 5 year
LISA mission. As seen, the ejected mergers drift into the LISA band with
eccentricities of $\sim 0.01$. The 2-body mergers appear in the LISA band with
eccentricities of nearly unity and leave with eccentricities above $\sim
10^{-3}$. The 3-body mergers appear outside of the LISA band.
}
\label{Fig:eof}
\end{figure}

We first consider the GW-driven eccentricity evolution of a representative BBH
from each population. Figure \ref{Fig:eof} shows the binary eccentricity
evolution for each of the three populations: ejected (blue), 2-body (green),
and 3-body (red). Each assumes purely GW-driven orbital evolution
\citep{Peters64}, and the population average initial orbital frequencies and
eccentricities ($<e_0>$, $<\fint>$) supplied by Paper I. The solid lines track
eccentricity vs. the rest-frame orbital frequency ($f_{\orb}$) while the
dashed lines track eccentricity vs. the peak rest-frame \textit{gravitational
wave} frequency, which we approximate as\footnote{This is simply the orbital
frequency if the binary had semi-major axis equal to its pericentre distance.
This is very similar to other, slightly more complicated, approximations used
in the literature \citep[\textit{e.g.},][]{Wen:2003}.}
\begin{equation}
f^{\rm{peak}}_{r} \approx 2f_{\orb}(1-e)^{-3/2}.
\label{Eq:fpeak}
\end{equation}
The filled circles denote the population average $<e_0>$ and $<\fint>$ at
formation. BBHs from each formation channel effectively form with the same
initial orbital frequency, corresponding approximately to the semi-major axis
below which a subsequent binary-single interaction would eject the BBH from
the cluster. The differences in the eccentricity evolution between each BBH
population is due primarily to the different initial eccentricities of each
population, which is regulated by the merger timescale for each formation
channel. Because binaries on highly eccentric orbits emit most of their GW
power in high harmonics of the orbital frequency (see \S \ref{S:res}),
differences in initial eccentricity results in the very different initial peak
GW frequencies denoted by the stars in Figure \ref{Fig:eof}.

To see how the characteristic peak frequency at formation, of a given
population with merger time scale $\mathcal{T}$, scales with the BBH orbital
parameters, one can combine the expression for the GW inspiral lifetime
$\mathcal{T} \approx t_{\rm c}(1-e^2)^{7/2}$ (see, \textit{e.g.}, \cite{Peters64}), where
$t_{\rm c}$ denotes the circular-orbit lifetime and $e$ is the eccentricity,
with the relation for $f^{\rm{peak}}_{r}$ given by Eq. (\ref{Eq:fpeak}), to
find,
\begin{equation}
f^{\rm{peak}}_{r,0}(\mathcal{T}) \approx 2\cdot10^{-5}\ \text{Hz}\ \left(\frac{\mathcal{T}}{10^{10}\text{yrs}}\right)^{-3/7} \left(\frac{a}{0.5\text{au}}\right)^{3/14} \left(\frac{m}{30M_{\odot}}\right)^{-11/14}.
\label{Eq:ftau}
\end{equation}
The characteristic time scales for our three considered populations, ejected
mergers, 2-body mergers, and 3-body mergers, are the Hubble time ($t_{\rm H}
\sim 10^{10}$ yrs), the binary-single encounter time ($\sim 10^{7}$ yrs), and
the BBH orbital time ($\sim 0.1$ yr), respectively. Substituting these time
limits into Eq. (\ref{Eq:ftau}), one finds that the three corresponding
characteristic peak frequencies are $\log{f^{\rm{peak}}_{r,0}} \approx -4.5$
(ejected BBH mergers), $\log{f^{\rm{peak}}_{r,0}} \approx -3.5$ (2-body BBH
mergers), and $\log{f^{\rm{peak}}_{r,0}} \approx 0$ (3-body BBH mergers).
These frequencies roughly correspond to average initial eccentricities for
each population of  $<e_0> \approx 1-10^{-1}$ (ejected),  $<e_0> \approx 1 -
10^{-2}$, (2-body), and $<e_0> \approx 1 - 10^{-5}$ (3-body). In other words,
the three different populations that we find to have very different
eccentricities in LIGO and LISA, are regulated simply by the time scale for
each to undergo a merger. More detailed numerical simulations will undoubtedly
find slight changes to the shape of the distributions, but the locations of
$f^{\rm{peak}}_{r,0}$ are robustly set by the three listed time scales, each
of which can be straight forwardly estimated\footnote{The least known time
scale is the encounter time scale, which is set by the number density of
single BHs in the GC core, $n_{\rm s}$; a number that at present is highly
uncertain. However, as described in \S \ref{S:GCdep}, both the location of the
2-body GW peak frequency and its normalization depends only weakly on $n_{\rm
s}$, and an order-of-magnitude uncertainty in $n_{\rm s}$ does therefore not
impact our results significantly.}.

As the binaries evolve towards higher orbital frequencies due to gravitational
radiation reaction, eccentricity is damped and the dashed lines in Figure
\ref{Fig:eof} approach a value that is a factor of two greater in frequency
than the solid lines. This is the usual factor of two relating the Keplerian
orbital frequency and the GW frequency for binaries on circular orbits;
binaries on circular orbits emit all of their GW power in the $n=2$ harmonic.

Figure \ref{Fig:eof} also shows the LISA sensitivity from
\cite{CornishLISASens:2018} shaded as a function of GW frequency in orange,
dark orange being the most sensitive. Also plotted are horizontal dotted
lines, above which LISA can measure the binary eccentricity. For $e>0.01$,
LISA is expected to measure a non-zero eccentricity for all resolved BBHs,
while for $e\sim0.001$, LISA should detect non-zero eccentricity for $\sim
90\%$ of identified BBHs over a 5 year mission \citep{Nishizawa+2016,
Seto:2016}. The black-dotted vertical line denotes the peak GW frequency at
which the ejected and 2-body populations will have approximately a 5 year
lifetime until merger, at $z=0.1$ (see Eq. \ref{Eq:fSNmax}). Hence, only the
BBHs with frequencies to the right of the black-dotted line (when a 5 year
LISA mission begins) are accessible to both LISA and LIGO.

The ejected BBHs enter the LISA band with detectable eccentricity of
$e\sim0.01$, but circularize below $e\sim0.001$ by the time they reach the
frequency above which a BBH will merge during the LISA mission lifetime.
Hence, BBHs in the ejected population that will be detectable by LISA and LIGO
have eccentricity that is only marginally measurable by LISA.

The majority of the 2-body BBHs form with peak GW power in the LISA band and
remain above $e=0.001$ for their entire voyage through the LISA band. Hence,
binaries from this population, which is of order $50\%$ of all BBH mergers
dynamically assembled in GCs (Paper I), will have eccentricity detectable by
LISA.

We note that \cite{BreivikRodLISA:2016} present a similar eccentricity
evolution plot in their Figure 1, in which they consider only tracks of BBHs
formed \emph{without including} relativistic effects (GW radiation in the
$N$-body equation-of-motion) considered in Paper I. We point out here that the
\emph{inclusion} of relativistic effects leads to the 2-body BBHs that are
absent from the plot in \cite{BreivikRodLISA:2016} and have eccentricity such
that they emit at peak GW frequencies directly in the LISA band (Paper I).
Although relativistic effects occur at higher order in the $N$-body equation-
of-motion for interactions in classical GCs, the 2-body mergers are not sub
dominant, as they constitute up to about $\sim 50\%$ of all the BBH cluster
mergers \citep{RodPND+2017, 2017arXiv171107452S, SamsingDOrazioLISAI:2018}.
This is a result of classical hardening, which gives each BBH repeated `tries'
for undergoing a merger inside the GC before ejection is possible
\citep[\textit{e.g.},][]{2017arXiv171107452S}.

Finally the 3-body BBHs are never prominent in the LISA band, rather, they are
so eccentric that they will form in the region of GW-frequency space between
LISA and LIGO \citep[\textit{e.g.},][]{ChenPau:2017, SamsingDOrazio:2018a}, and enter
the LIGO band with $e\sim0.1$ \citep[\textit{e.g.},][]{2017arXiv171107452S}.
These highly eccentric BBHs are the relevant population for the work by
\cite{ChenPau:2017, SamsingDOrazio:2018a}, who point out that very high
eccentricity BBHs will have strain that is too low to be detected by LISA, but
would be delectable by proposed GW detectors such as DECIGO \citep{DECIGO} and
Tian Qin \citep{TianQin} in the $\sim1-10$~Hz band.

We now compute the tracks of binaries from each population in GW amplitude vs.
GW frequency through the LISA band down to merger. This will tell us when the
BBHs are detectable and the relevant eccentricities to which LISA and LIGO
have access for each population.

%%%%%%%%%%%%%%%%%%%%%%%%%%%%%%%%%%%%%%%%%%%%%%%%
%%% FIGURE %%%
%%%%%%%%%%%%%%%%%%%%%%%%%%%%%%%%%%%%%%%%%%%%%%%%    

\subsection{Gravitational wave amplitudes}
\label{S:res}

We have shown that the three BBH populations considered here exhibit an
interesting range of eccentricities in the LISA and LIGO bands. We now wish to
show that BBHs with these eccentricities and orbital frequencies will be
detectable as individually resolved sources by LISA or LIGO. To do this we
compute the characteristic strain of a representative BBH from each of the
populations discussed above. We also compute the signal-to-noise ratio ($S/N$)
of each eccentric BBH population in the LISA or LIGO band.

To compute the characteristic strain vs. GW frequency, $h_c(f)$, during the
orbital evolution of an eccentric binary, we must consider the wide range of
harmonics at which an eccentric binary emits gravitational waves. The
$n^{\rm{th}}$ harmonic is emitted at a rest-frame GW frequency of $f_r = n
f_{\orb}$, where $f_{\orb}$ is the rest frame orbital frequency of the binary
and the rest-frame and observed GW frequency are related through the
cosmological redshift $z$, $f_r = (1+z)f$. For a binary on a circular orbit,
all power is emitted at the $n=2$ harmonic, but in general, the characteristic
strain in GWs at the $n^{\rm{th}}$ harmonic of the orbital frequency is
\citep{BarackCutler:2004},
\begin{equation}
h_{c,n}(f) = \frac{1}{\pi D} \sqrt{\frac{2G}{c^3}\frac{dE_n}{df_r} },
\label{Eq:hc}
\end{equation}
where $D$ is the luminosity distance to the source. The energy emitted per GW
frequency at the $n^{\rm{th}}$ harmonic is given by
\citep{Enoki+2007,Huerta+2015, CSD:2017},
\begin{equation}
\frac{dE_n}{df_r} = \frac{(G\Mc)^{5/3} }{ 3 \pi^{1/3} (1+z)^{1/3}  f^{1/3}  } \left(\frac{2}{n}\right)^{2/3} \frac{g(n, e)}{F(e)},
\label{Eq:dEndf}
\end{equation}
where $\Mc \equiv M q^{3/5}/(1+q)^{6/5}$ is the chirp mass for a binary with
total mass $M=M_1+M_2$ and mass ratio $q\equiv M_2/M_1 \leq1$. The functions
$g(n,e)$ and $F(e)$ are defined in \cite{PetersMatthews:1963}.

The above expression requires evaluation of the eccentricity at all points
along the binary evolution. This is found by solving
\citep[\textit{e.g.,}][]{Enoki+2007, Huerta+2015},
\begin{equation}
\frac{f_{\orb}}{f_{0}} = \left[\frac{1-e^2_0}{1-e^2} \left(\frac{e}{e_0}\right)^{12/19} \left(\frac{1+\frac{121}{304}e^2}{1+\frac{121}{304}e^2_0} \right)^{870/2299} \right]^{-3/2},
\label{Eq:eof}
\end{equation}
at a specified orbital frequency, given the initial binary orbital frequency
and eccentricity.

Because we are considering BBH sources with lifetimes that can be much longer
than the LISA mission lifetime, we must adjust the characteristic strain to
take into account this finite observation lifetime in the LISA band, $\tau$.
This is primarily because the characteristic strain is a measure of the signal
per frequency bin, and in the case of BBH sources, one must take into account
the amount of time that a BBH spends in a given frequency bin compared to the
total observation time.  To do so, observe that the energy produced by a GW
source at a given frequency is proportional to the squared GW strain $h^2$
times the number of cycles emitted at that frequency. Hence for comparing to
the LISA/LIGO sensitivity curves, we compute the characteristic strain,
$h_c(f) \equiv h(f)\sqrt{N} \propto h(f) \sqrt{f^2/\dot{f}}$, where $\dot{f}$
is the rate of change of the binary frequency, making $N$ the number of cycles
spent in the frequency interval $\Delta f$ \citep[][our Eq.
\ref{Eq:hc}]{SesanaHMV:2005}. When the number of cycles per frequency interval
is larger than the total number of cycles observed at that frequency, in an
observation time $\tau$, the value of $N$ in the characteristic strain becomes
simply $f\tau$, and the total signal is set by how long LISA can gather signal
in that frequency bin. In Figures \ref{Fig:hctracks} and \ref{Fig:hctracks2}
below, the characteristic strain is computed to take into account a finite
observation time $\tau=5$~years unless otherwise specified.

Each panel in Figure \ref{Fig:hctracks} shows the full GW emission from
eccentric BBHs over their entire lifetimes. From left to right, we plot a
representative BBH track from the ejected, 2-body, and 3-body BBH populations
using the same mean initial orbital frequency and eccentricity as in Figure
\ref{Fig:eof}. In each panel, the solid black and grey lines are the
sensitivity curves for the LISA and LIGO instruments respectively. To
elucidate the binary evolution, we draw thick dashed lines for each population
that trace the GW frequency of peak GW emission. The thin solid lines of a
given color (blue, green, red) represent the GW-emission at a fixed binary
orbital frequency, at a snapshot in the binary's evolution.

The `knee' in the BBH evolution tracks for the ejected and 2-body BBHs,
occurring at $f \sim 10^{-1.8}$~Hz (where the thin and thick dashed lines
meet), is the transition between observation-duration limited and 
binary-inspiral limited characteristic strain. Practically, BBHs emitting at peak GW
frequencies to the right of this knee, when LISA turns on, will merge during
the mission lifetime, while those with starting frequency low of the knee will
have negligible motion through frequency space as seen in Figure
\ref{Fig:hctracks}. A thin dashed line is drawn for reference to show the
evolution of the characteristic strain over an infinite observation time.

Early in the evolution, the GW emission is much lower than what is expected
for a circular-orbit binary at the same orbital frequency. This is because the
binary is emitting a large spread of GW frequencies above its orbital
frequency, resulting in the GW emission profile traced out by the solid
colored lines. As the binary orbital frequency and eccentricity evolve, this
profile becomes narrower, and higher in amplitude. In the early stage of
evolution, when the binary is still very eccentric, the frequency of peak GW
emission changes very little while the amplitude of emission increases
rapidly. Eventually, the binary circularizes and the GW emission profile
approaches a delta-function with peak at GW frequency tracking twice the
binary orbital frequency.

The thick dashed lines tracing peak GW emission approximately follow the spine
of the time series of constant $f_{\orb}$ emission profiles. At
circularization, each track joins the $f^{-1/6}$ power law evolution expected
for a BBH on a circular orbit (this power law would be extended to low
frequencies for a binary initially on a circular orbit). This coincides with
the point where the dashed and solid lines become parallel in Figure
\ref{Fig:eof}. The cutoff in characteristic strain at high frequencies
signifies merger, where the orbital period is above the maximum plunge
frequency \citep{BarackCutler:2004}.

Each line of constant orbital frequency is made up from the contribution of
GWs at many harmonics of the orbital frequency; the higher the eccentricity,
the more harmonics contribute. To show this, the lighter colored markers
connected by a thin line in each panel of Figure \ref{Fig:hctracks} denote the
evolution of three selected harmonics of the GW emission. The `$*$'s pick out
the $n=2$ harmonic over the course of the binary evolution, while the `$x$'s
and filled circles represent higher harmonics as labeled. It is clear that the
$n=2$ harmonic rises to dominate over the higher order harmonics as the binary
circularizes.

From left to right, the effect of a larger initial eccentricity can be readily
observed. The middle and right panels exhibit a longer low frequency tail of
GW emission early in the binary evolution. The higher eccentricity cases also
sit at a nearly constant, and higher peak GW frequency for a larger range of
characteristic amplitudes before circularizing, at which point the GW emission
follows the orbital frequency to higher values and merger. In the right panel,
the initial eccentricity is so high that the solid lines of constant orbital
frequency near formation cannot be completed with the $100,000$ harmonics
included in this rendering. This manifests in the incomplete solid red lines
at $\sim1$~Hz and below $h_c\sim10^{-25}$.

It is interesting to note that if the representative 3-body BBH studied here
is at least twice as massive and within 10 Mpc, or instead consists of
intermediate mass BHs at cosmological distance, the spread of emitted
GWs, at a single point in time, spans both LISA and LIGO simultaneously. Such
simultaneous multiband detection was pointed out by \citep{KocsisLevin:2012},
who have computed similar eccentric GW emission tracks for a different
dynamical formation scenario.

%%%%%%%%%%%%%%%%%%%%%%%%%%%%%%%%%%%%%%%%%%%%%%%%
%%% FIGURE %%%
%%%%%%%%%%%%%%%%%%%%%%%%%%%%%%%%%%%%%%%%%%%%%%%%    
\begin{figure*}
\begin{center}$
\begin{array}{ccc}
\includegraphics[scale=0.354]{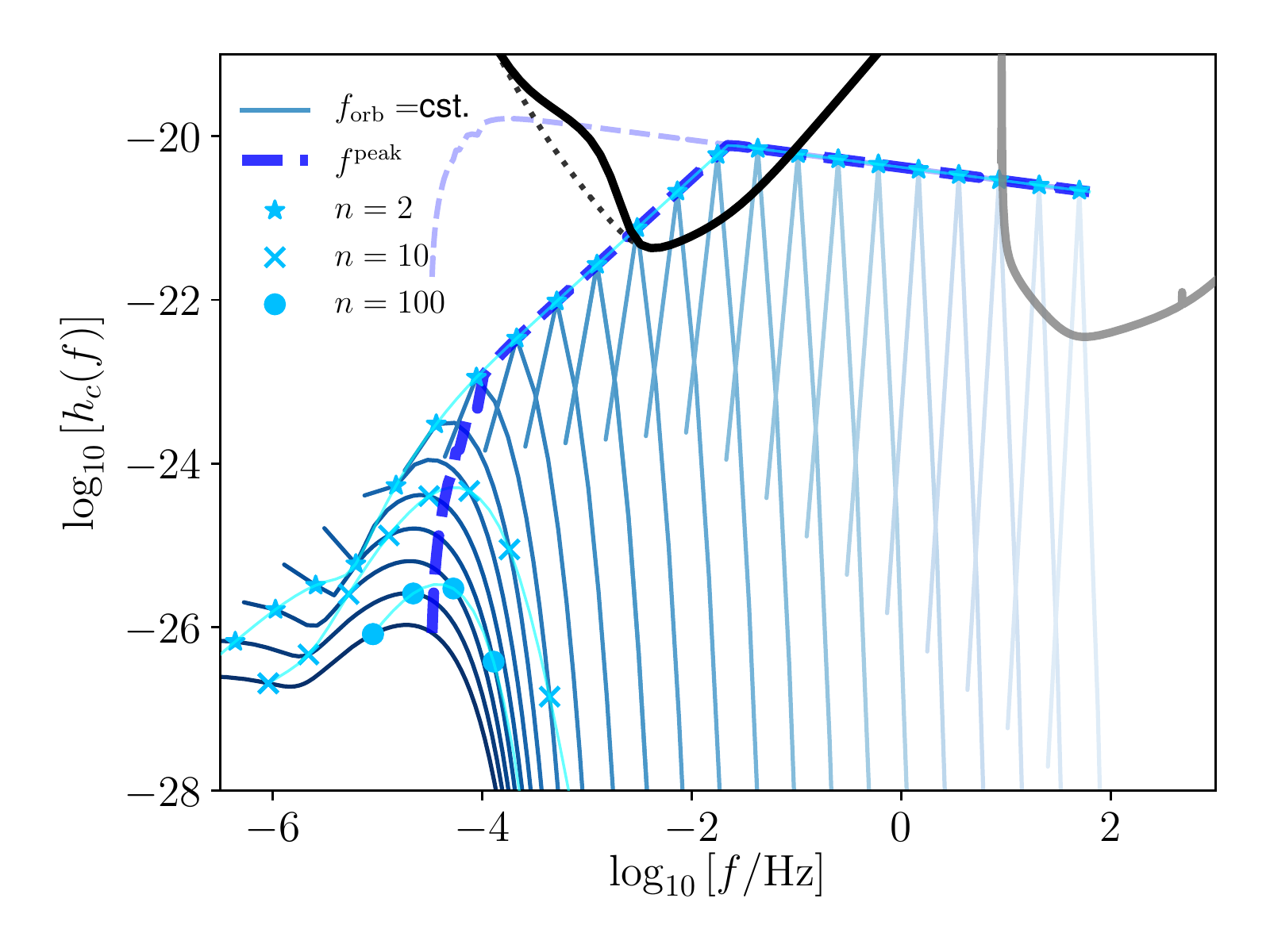} &
\includegraphics[scale=0.354]{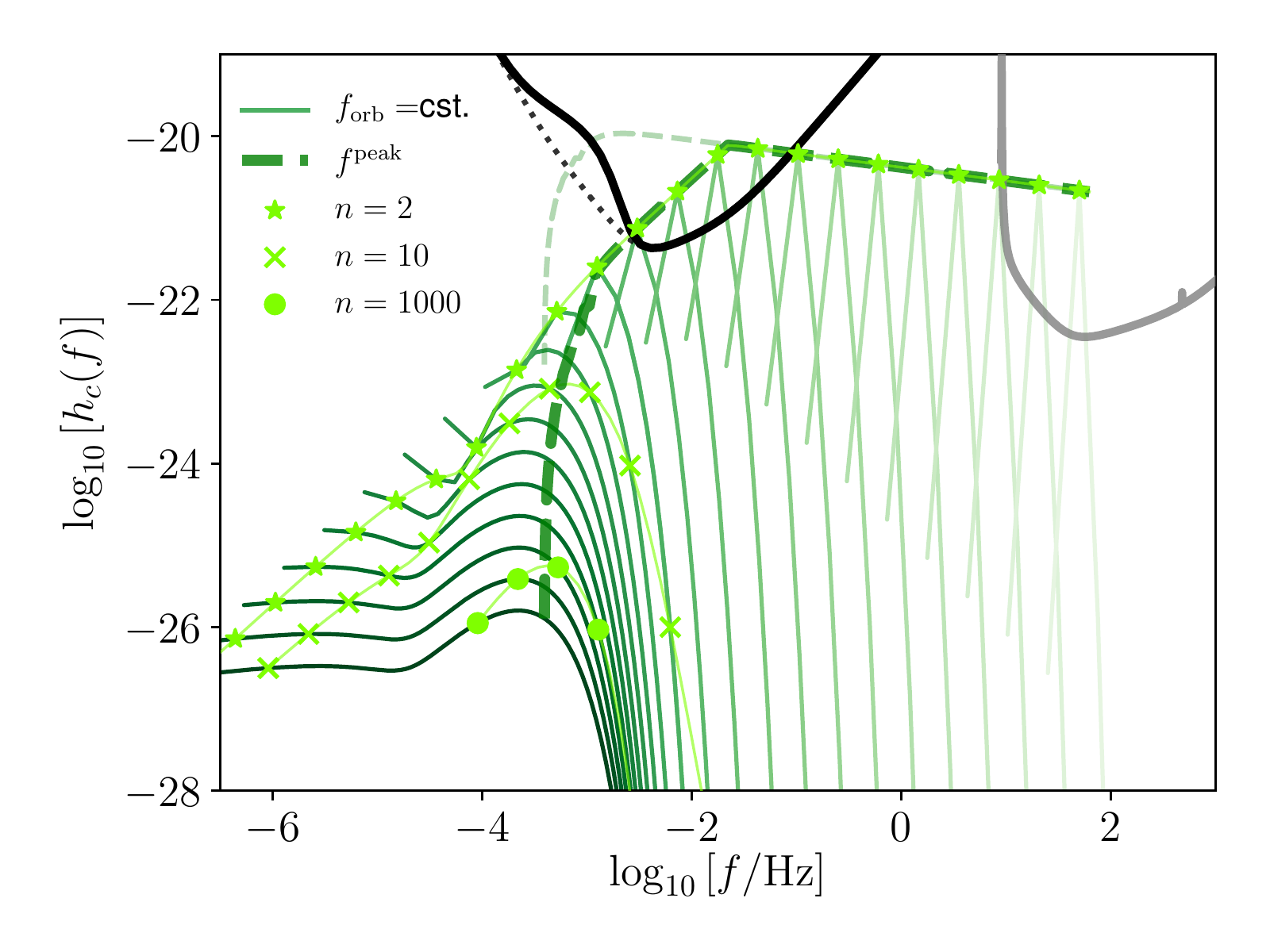} &
\includegraphics[scale=0.354]{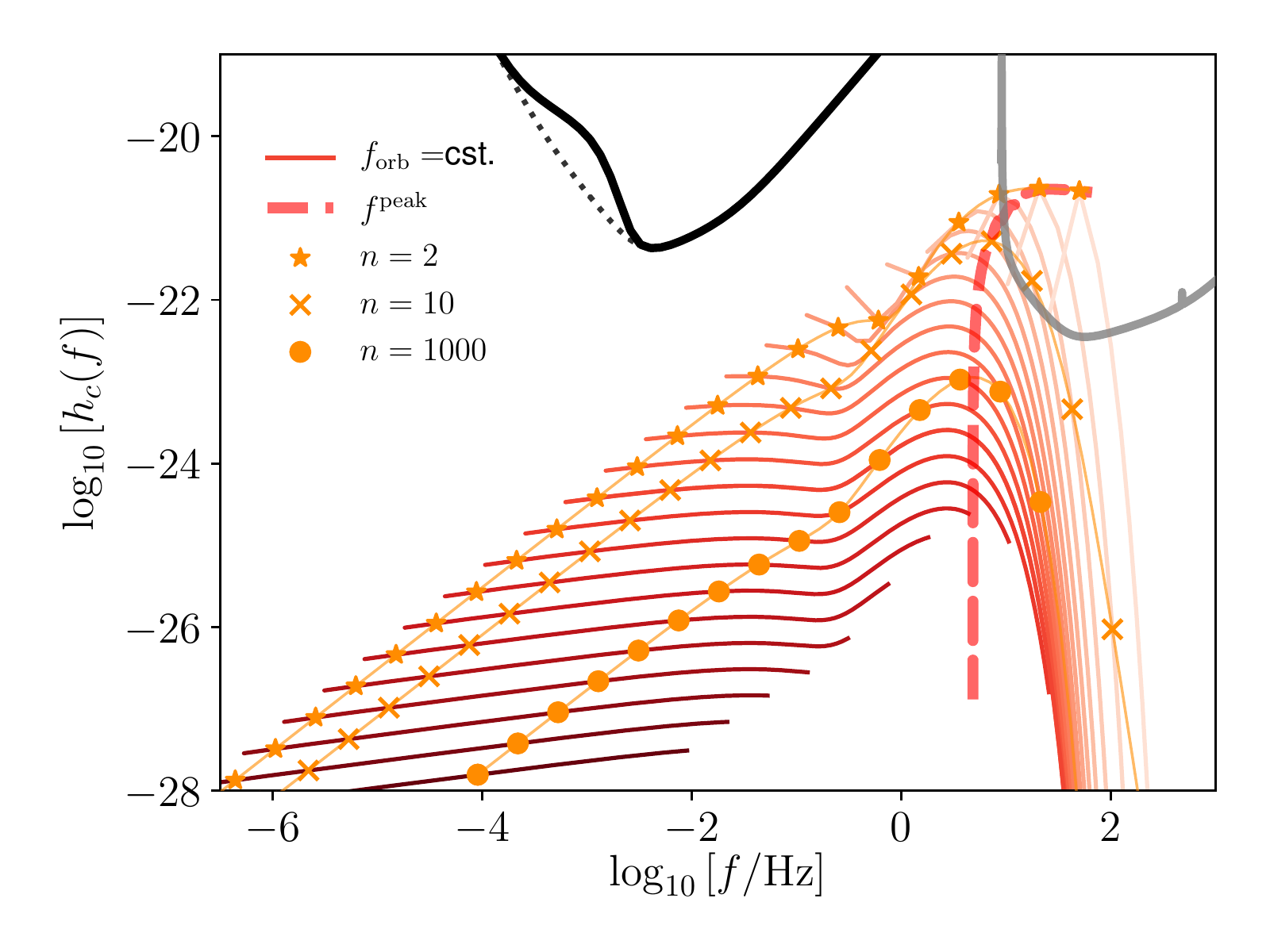} 
\end{array}$
\end{center}
\vspace{-15pt}
\caption{
The tracks of eccentric $30+30\Msun$ BBHs through GW frequency space. The BBHs
are assumed to be at a distance of 500 Mpc. The three panels show tracks for
each of the three populations discussed in the text; ejected (left), 2-body
(center), and 3-body (right). The thick black line is the LISA instrument
sensitivity curve with (solid) and without (dotted) the galactic binary
background \citep{CornishLISASens:2018}. The grey line is the advanced LIGO
design-sensitivity curve \citep{aLIGO:2015}. For each population (each color set /
panel) the solid lines represent the GW emission at a specific binary orbital
frequency, at one slice in binary-rest-frame time. The lines of constant
orbital frequency are shaded dark to light for early to late time evolution.
Early in the evolution, the binary is eccentric and emits at a spread of GW
frequencies, with the most power at frequency $f^{\rm{peak}}$; evolution at
$f^{\rm{peak}}$ is drawn as a thick dashed line. As the binary circularizes,
it emits at predominately the $n=2$ harmonic which is the power law that the
dashed line reaches at high GW frequencies. The thin dashed lines show the
characteristic strain for an infinite observation time (see text). The over-
plotted, brighter markers track only single harmonics of the binary evolution.
The `$*$'s track the $n=2$ harmonic that dominates upon circularization, the
$x$'s and the filled circles denote higher order harmonics as labeled. We
compute the first 100,000 harmonics in order to draw the lines of constant
orbital frequency.
}
\label{Fig:hctracks}
\end{figure*}

%%%%%%%%%%%%%%%%%%%%%%%%%%%%%%%%%%%%%%%%%%%%%%%%
%%% FIGURE %%%
%%%%%%%%%%%%%%%%%%%%%%%%%%%%%%%%%%%%%%%%%%%%%%%%    
\begin{figure}
\begin{center}$
\begin{array}{c}
\includegraphics[scale=0.23]{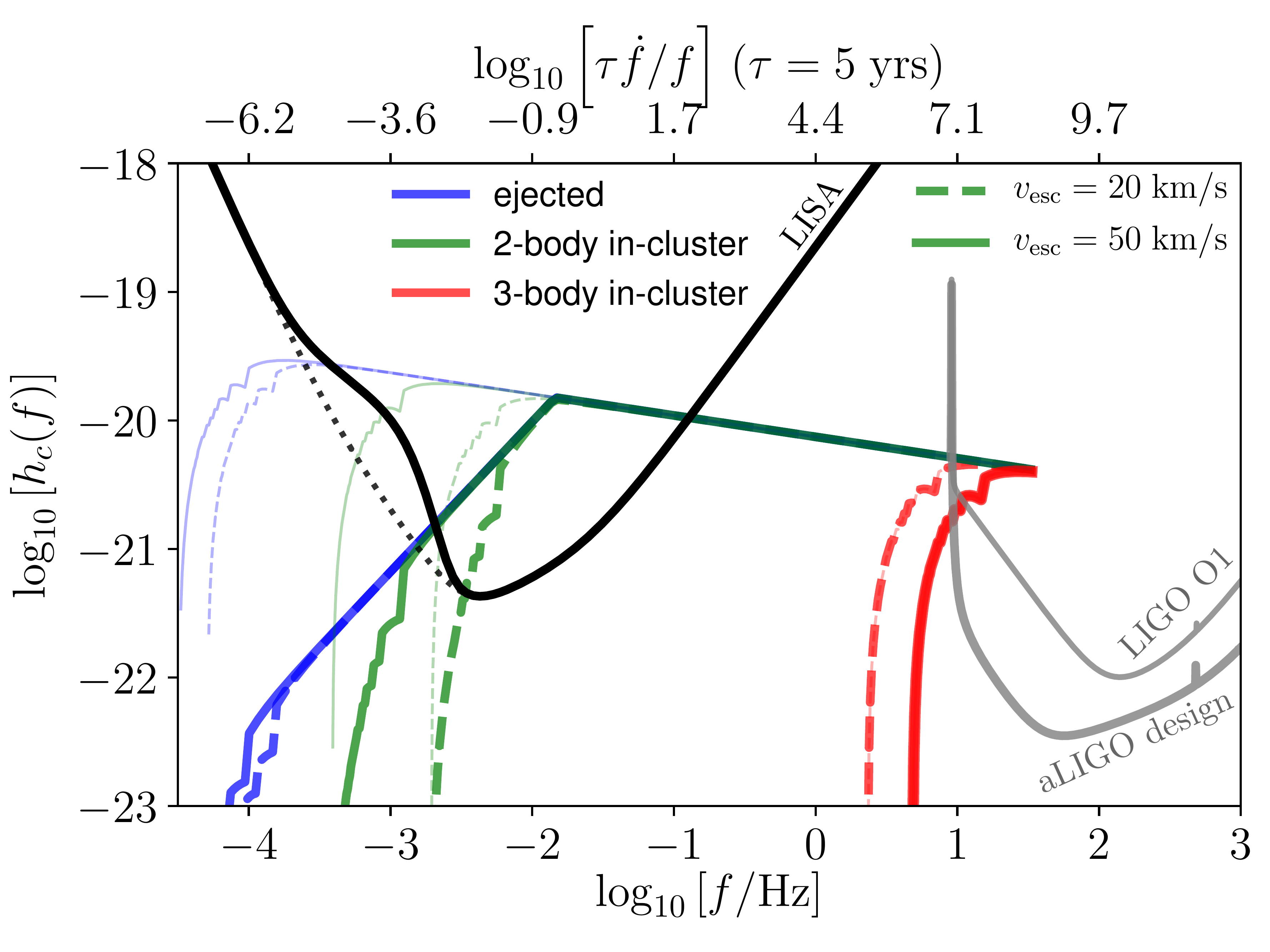} \\
\end{array}$
\end{center}
\vspace{-15pt}
\caption{
Tracking only the peak GW frequency of each population (thick dashed lines in
Fig. \ref{Fig:hctracks}), assuming a BBH distance of 500 Mpc. Solid lines
denote representative BBHs coming from a globular cluster with an escape
velocity of $v_{\rm{esc}}=50$~km/s. Dashed lines assume
$v_{\rm{esc}}=20$~km/s. Different globular cluster escape velocities result in
a different initial orbital frequency and eccentricity distribution for a
given population. On the top x-axis, we show the ratio of the number of binary
orbits that a binary will undergo in a LISA mission lifetime of $\tau = 5$
years divided by the number of orbits needed for the binary to evolve by of
order its starting frequency. Labeled sensitivity curves are plotted for
reference.
}
\label{Fig:hctracks2}
\end{figure}

In Figure \ref{Fig:hctracks2} we summarize the tracks of GW emission of the
three BBH populations, plotting only the evolution of the peak harmonic,
$h_c(f^{\rm{peak}})$. The thick lines show the peak characteristic
strain assuming a LISA lifetime of 5 years, while, for reference, the thin
lines show the peak characteristic strain assuming an infinite observation
time. The thin lines are the same peak harmonic tracks as those plotted
in Figure \ref{Fig:hctracks}, for the fiducial GC parameters
($v_{\rm{esc}}=50$~km s$^{-1}$). To show the dependence on GC properties, the
dashed lines show the tracks of representative BBHs from each population
coming instead from a GC with $v_{\rm{esc}}=20$~km s$^{-1}$. Dependence on GC
properties is discussed below in \S \ref{S:GCdep}.

On the top x-axis of Figure \ref{Fig:hctracks2} we quantify the length of a
the binary track through GW frequency space over the course of a 5 year LISA
mission. For a figure of merit we use the number of cycles observed at
frequency $f$ in the observing time $\tau$ divided by the approximate number
of cycles that a binary spends at frequency $f$, $f^2/\dot{f}$ \citep[see
\textit{e.g.},][]{SesanaHMV:2005}, where $\dot{f}$ is the rate of change of
binary orbital frequency due to GW decay. The binary frequency changes by of
order the starting value in a 5 year window at approximately the location of
the knee feature for the ejected and 2-body BBH populations. BBHs to the left
of the knee can still be detected by LISA, but will change in frequency only
by the small amount indicated by the top x-axis. While these systems will not
merge during the LISA lifetime, they still will have measurable orbital
parameters, and might still merge while a future ground-based instrument is
operating.

\subsubsection{Signal-to-noise ratios}

The tracks of characteristic strain in Figures \ref{Fig:hctracks} and
\ref{Fig:hctracks2} give a by-eye estimate of the $S/N$ for
the drawn LISA and LIGO sensitivity curves \citep[see
\textit{e.g.},][]{CornishLISASens:2018, MooreColeBerry:2015}.  Here we
quantify this, computing the $S/N$ in the LISA and LIGO bands as,
\begin{equation}
\left( \frac{S}{N} \right)^2 \approx 2 \sum^{n_{\max}}_{n=1} \int^{f(\tau)}_{ f_{\rm{start}} } \frac{\mathcal{F}(f) h^2_{c,n}(f)}{ f  P_n(f)  } \frac{df}{f},
\label{Eq:SNR}
\end{equation}
where the summation is over the relevant harmonics for an eccentric orbit
\citep[\textit{e.g.},][]{OlearyKocsis:2009}. The lower limit of integration
$f_{\rm{start}} = n f_{{\rm orb}}(0)/(1+z)$ is the GW frequency being emitted
at the $n^{\rm th}$ harmonic of the orbital frequency when LISA or LIGO begin
observation, and $f(\tau)$ is either the GW frequency at merger, or the GW
frequency in the $n^{\rm{th}}$ harmonic of the orbital frequency to which the
binary evolves after observation time $\tau$. $P_n(f)$ is the noise power
spectral density of the detector and $\mathcal{F}(f)$ is the 
detector-dependent sky and polarization averaged signal response function.  For an
interferometer with perpendicular arms like LIGO, $\mathcal{F}(f)=1/5$. For
LISA, $\mathcal{F}(f)$ is generally frequency dependent and is given as a
fitting formula in \cite{CornishLISASens:2018}. We note that, following
convention, the sensitivity curves drawn here for LIGO are simply $\sqrt{f
P_n(f)}$, whereas those drawn here for LISA are $\sqrt{f
P_n(f)/\mathcal{F}(f)}$.

To find $f(\tau)$ we solve the coupled ordinary differential equations for
$de/dt$ and $d f_{\orb}/dt$. We first evolve a BBH from its initial orbital
frequency and eccentricity to the orbital frequency and eccentricity
corresponding to the GW frequency $f_{\rm{start}}$, at the beginning of
observation. We then evolve the binary parameters for a rest-frame lifetime of
$\tau/(1+z) = 5/(1+z)$~yrs and truncate at merger if necessary.

\begin{figure}
\begin{center}
\includegraphics[scale=0.5]{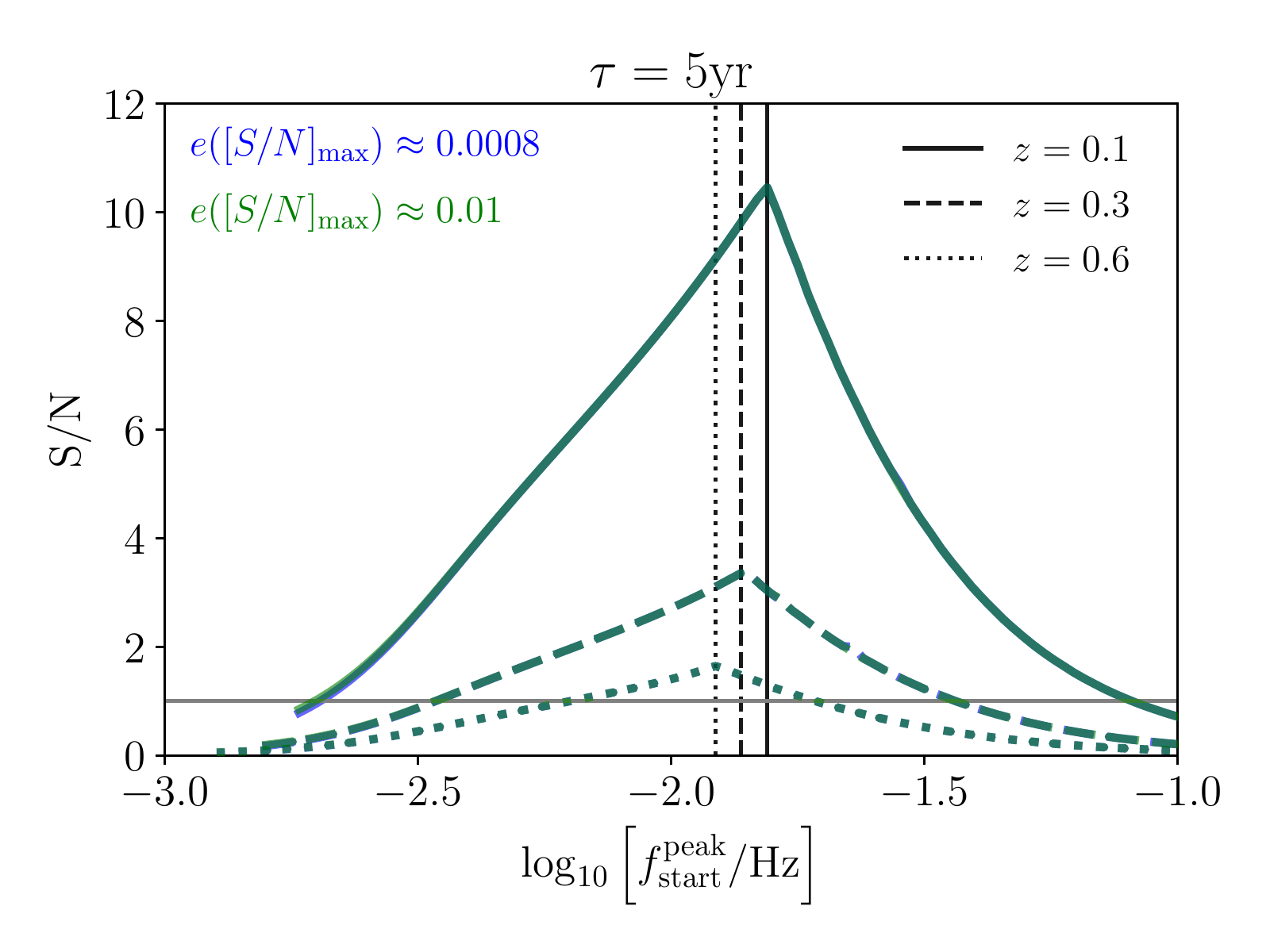}
\end{center}
\vspace{-15pt}
\caption{
LISA signal-to-noise ratio for BBHs from the ejected (blue) and 2-body (green)
populations, at different redshifts (the $S/N$ is nearly identical for both
populations). The x axis varies the peak GW frequency being emitted by the
binary at the beginning of LISA observations, $f^{\rm{peak}}_{\rm{start}}$.
The black vertical lines are drawn at the position of maximum $S/N$ for each
redshift. Labeled in the same colors as the corresponding BBH population are
eccentricities at the value of $f_{\rm{start}}$ where the $S/N$ is maximum in
the LISA band (not greatly changing over the redshift range considered here).
The black lines also indicate the GW peak frequency above which the BBHs in
the ejected and 2-body populations will merge during the assumed 5 year LISA
mission.
}
\label{Fig:SNR}
\end{figure}

Figure \ref{Fig:SNR} presents the LISA $S/N$ for both the ejected and 2-body
populations as a function of the starting $30+30 \Msun$ BBH GW frequency when
LISA turns on. We consider three different redshifts to the source, our
fiducial $z=0.1$ (500 Mpc), $z=0.3$, and $z=0.6$. The shape of this curve can
be understood partly from the shape of the LISA sensitivity curve, decreased
sensitivity for frequencies departing from $\approx10^{-2.5}$~Hz, and partly
from the  observation-time dependence of the characteristic strain.

For BBHs at starting frequencies to the left of the peak $S/N$ in Figure
\ref{Fig:SNR}, the change in binary frequency over the LISA lifetime $\tau$ is
approximately zero (see the top x-axis of Figure \ref{Fig:hctracks2}). Then
the integral over $f$ can be replaced by multiplication by $\Delta f$ and the
$S/N$ becomes
\begin{equation}
\left( \frac{S}{N} \right)^2 \propto \frac{\mathcal{F}(f) h^2(f)}{P_n(f)} \frac{f^2}{\dot{f}} \frac{\Delta f}{f} \approx \frac{\mathcal{F}(f) h^2(f)}{P_n(f)} f \tau,
\end{equation}
which increases with increasing GW frequency as more cycles can be accumulated
in an observation time. For BBHs at starting frequencies to the right of the
peak $S/N$ the binary now evolves significantly in a LISA observation time and
the $S/N$ drops with increasing frequency because GW-driven BBHs spend fewer
cycles at higher frequencies. We note that the $S/N$ calculations presented in
Figure \ref{Fig:SNR} are relevant for source detection only. For the
measurement of eccentricity, a fisher matrix approach has been applied in
\citet{Nishizawa+2016} and \citet{Gondan+2018}. Furthermore,
\citet{Nishizawa+2017} has considered the ability to select formation models
in addition to parameter estimation.

The frequency of maximum $S/N$ occurs approximately where the inspiral
time of a BBH equals the mission lifetime. Hence, we estimate the binary
orbital eccentricity at this frequency as it signifies the highest $S/N$
systems, and also delineates between BBHs that would be observable over the
duration of a joint LISA and ground-based mission, and those that stay in the
LISA band. The peak $S/N$ occurs at
\begin{equation}
f([S/N]_{\max}) \approx 10^{-1.8} \rm{Hz} \ \left(\frac{\tau}{5 \rm{yr}}\right)^{-3/8} \left(\frac{\Mc}{26} \frac{1+z}{1+0.1}\right)^{-5/8},
\label{Eq:fSNmax}
\end{equation}
where we have assumed that $e \ll 1$ at the $S/N$ peak location. At the 
rest-frame orbital frequency corresponding to this peak GW frequency, the binary
orbital eccentricities at peak $S/N$ are labeled for each population in Figure
\ref{Fig:SNR}. We find that, using the average initial orbital frequencies and
eccentricities of each population, the average eccentricities at peak $S/N$
are $e([S/N]_{\max}) \approx 8 \times 10^{-4}$ for the ejected population and
$e([S/N]_{\max}) \approx 0.01$ for the 2-body BBHs, with weak redshift
dependence for $z\lsim 1$.

We draw a vertical dotted line in Figure \ref{Fig:eof} to indicate the
frequency of the peak $S/N$ for a fiducial redshift of $z=0.1$. From this we
can visually track the BBH evolution from detection to when it leaves the LISA
band. By the time the representative ejected BBH leaves the LISA band at a GW
frequency of $\sim0.1$~Hz, its eccentricity will decay to a $\sim10$ times
smaller value of $e\sim 10^{-4}$. The representative 2-body BBHs, exit the
LISA band with an eccentricity of $e\sim 10^{-3}$. Hence, based on estimates
for LISA's ability to measure eccentricity \citep{Nishizawa+2016,
Nishizawa+2017, Seto:2016}, we find that the eccentricity of BBHs from the
2-body population is measurable by LISA for the entire time that they are
detectable in the LISA band. In contrast, the ejected BBH population has
marginally measurable eccentricity only at the earliest point in its
detectable evolution through the LISA band.

Figure \ref{Fig:SNR} shows not only that the ejected and 2-body BBHs can be
detected by LISA, but also that they will have similar detection probabilities
given that they have nearly the same expected $S/N$ in LISA. This is important
for inferring the fraction of each in an observationally unbiased manner from
eccentricity measurements \citep[see, \textit{e.g.},][]{Nishizawa+2017}.

For the 3-body population, the relevant $S/N$ measurement is in the LIGO band.
We compute the expected $S/N$ for the LIGO sensitivity that deteced BBHs so
far \citep[approximated by the `Early' curve from][]{LIGOObsDoc:2018}, and the
aLIGO design sensitivity \citep{aLIGO:2015}. For the current LIGO sensitivity
curve (denoted O1 in Figure \ref{Fig:hctracks2}) we find $S/N\sim5$ for both
the 3-body BBH mergers (entering LIGO at $e\sim0.1$) and a BBH on a circular
orbit, each made of two $30\Msun$ BHs at $z=0.1$. Note that we use the sky and
polarization averaged $S/N$ to compute this value ($\mathcal{F}=1/5$ in Eq.
\ref{Eq:SNR}). For the aLIGO design sensitivity we find a large improvement;
the representative 3-body BBH will have $S/N = 40$ in the LIGO band,
approximately the same as its circular-orbit counterpart. The design
sensitivity aLIGO $S/N$ drops to a value of $8$ for the 3-body BBHs a redshift
of $z=0.32$. The prospects for measuring eccentricity for this population was
studied in \citet{2014PhRvD..90j3001T} and \citet{Gondan+2018}.

In this section we have shown that LISA can marginally detect the eccentricity
of the ejected BBHs for the first portion of their evolution in the LISA band
and importantly, that LISA can observe the eccentricity decay of the 2-body
BBHs over the $\lsim5$ years that they spends in the LISA band. After leaving
the LISA band, each BBH population will have a short $\sim 11 (1+z)$ day decay
to merger that will pass the BBH through the DECIGO and LIGO bands.

The 3-body BBH population is not observable in the LISA band (unless they are
very close so that lower harmonics pass the LISA $S/N$). As pointed out in
other recent works
\citep{RodPND+2017,SamsingDOrazio:2018a,SamsingDOrazioLISAI:2018}, however,
Figures \ref{Fig:hctracks} and \ref{Fig:hctracks2} and our calculation of the
$S/N$ conclude that the 3-body population is detectable by LIGO, and should
also be detectable as a highly eccentric source in a future DECIGO-like
mission \citep{ChenPau:2017, SamsingDOrazio:2018a}. Figure \ref{Fig:eof} shows
that this population also has an eccentricity $e\sim0.1$ upon entering the
LIGO band, which should be easily measurable \citep[\textit{e.g.},][]{Gondan+2018}. Such
a highly eccentric population in the LIGO band could also be created by the
Kozai-Lidov mechanism \citep{AntoniniPerets:2012, Hoang+2018,
2018ApJ...853...93R, 2018arXiv180205718R, 2017ApJ...836...39S} or GW capture
from single+single interactions \citep{OlearyKocsis:2009, KocsisLevin:2012,
Gondan+2017} in nuclear star clusters.  However, these formation channels may
have other unique properties that would allow future GW observations to
distinguish between the two, for example correlations between BBH chirp mass
and eccentricity. 

%Future work should identify any correlations such as those
%between BBH chirp mass and eccentricity in the GC-formed populations.

%%%%%%%%%%%%%%%%%%%%%%%%%%%%%%%%%%%%%%%%%%%%%%%%
%%% FIGURE %%%
%%%%%%%%%%%%%%%%%%%%%%%%%%%%%%%%%%%%%%%%%%%%%%%%    

\section{Stochastic gravitational wave background}
\label{S:GWB}

Apart from individually resolvable sources in the LISA band, there will be
many more binaries contributing weaker signals to a present day stochastic
background of gravitational waves, which manifests as a low amplitude
confusion noise for LISA.

The gravitational wave background (GWB) encodes information on an entire
population of GW sources \citep[see][for a review]{RosadoGWB:2011}. For binary
systems, this information depends not only on the number density and
distribution of BBH orbital parameters over cosmic time, but also the
mechanisms which drive the binaries together. For example, previous works have
considered the GWB of the most massive BBHs ($\gsim10^9\Msun$) at the ends of
their inspirals \citep[\textit{e.g.},][]{SVC:2008, SVV:2009, SesanaGWB:2013,
McOsPr:2014}, which is a prime target of the Pulsar Timing Arrays
\citep[PTAs][]{PTAs} at low frequencies ($10^{-9}$~Hz). These studies have
shown that the inclusion of gas driven orbital decay
\citep{KocsisSesana:2011}, stellar scattering \citep{RassMerritt:2017}, binary
stalling \citep{Dvorkin+2017}, significant binary eccentricity
\citep{Enoki+2007, Huerta+2015, CSD:2017}, or all of the above
\citep{KelleyGWB+2017} can have dramatic effects on the predicted GWB.

\cite{SesanaLISALIGO:2016} and \citet{Nishizawa+2016} have recently
shown that stellar mass BBHs will also contribute to a GWB that is most
significant in the LISA band. Here we extend this calculation to include the
eccentric stellar mass BBHs formed in GCs. Specifically, we compute the GWB
arising from the combination of the three, cluster-formed BBH populations of
Paper I and compare to a population of circular-orbit binaries representative
of those that may have formed in the field. We note that the following
analysis is restricted to the equal mass case, described in the introduction
to Section \ref{sec:Orbital Evolution and GW Emission}, whereas the
calculations in \cite{SesanaLISALIGO:2016} sample the BH masses from a BH mass
function. However, our main results reported below are generally valid and do
not directly follow from our simplified model.

\begin{figure*}
\begin{center}$
\begin{array}{cc}
\includegraphics[scale=0.5]{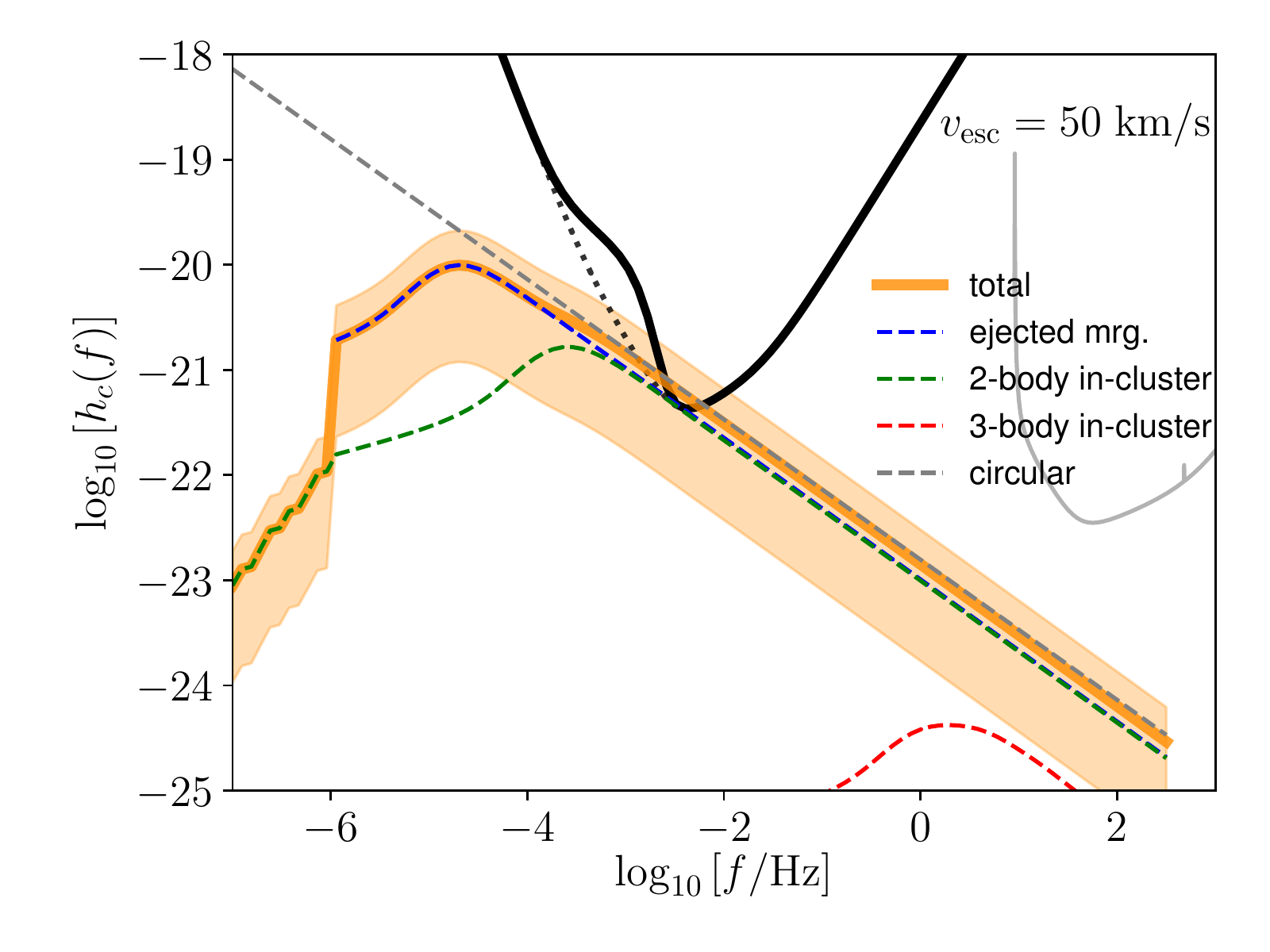} 
\includegraphics[scale=0.5]{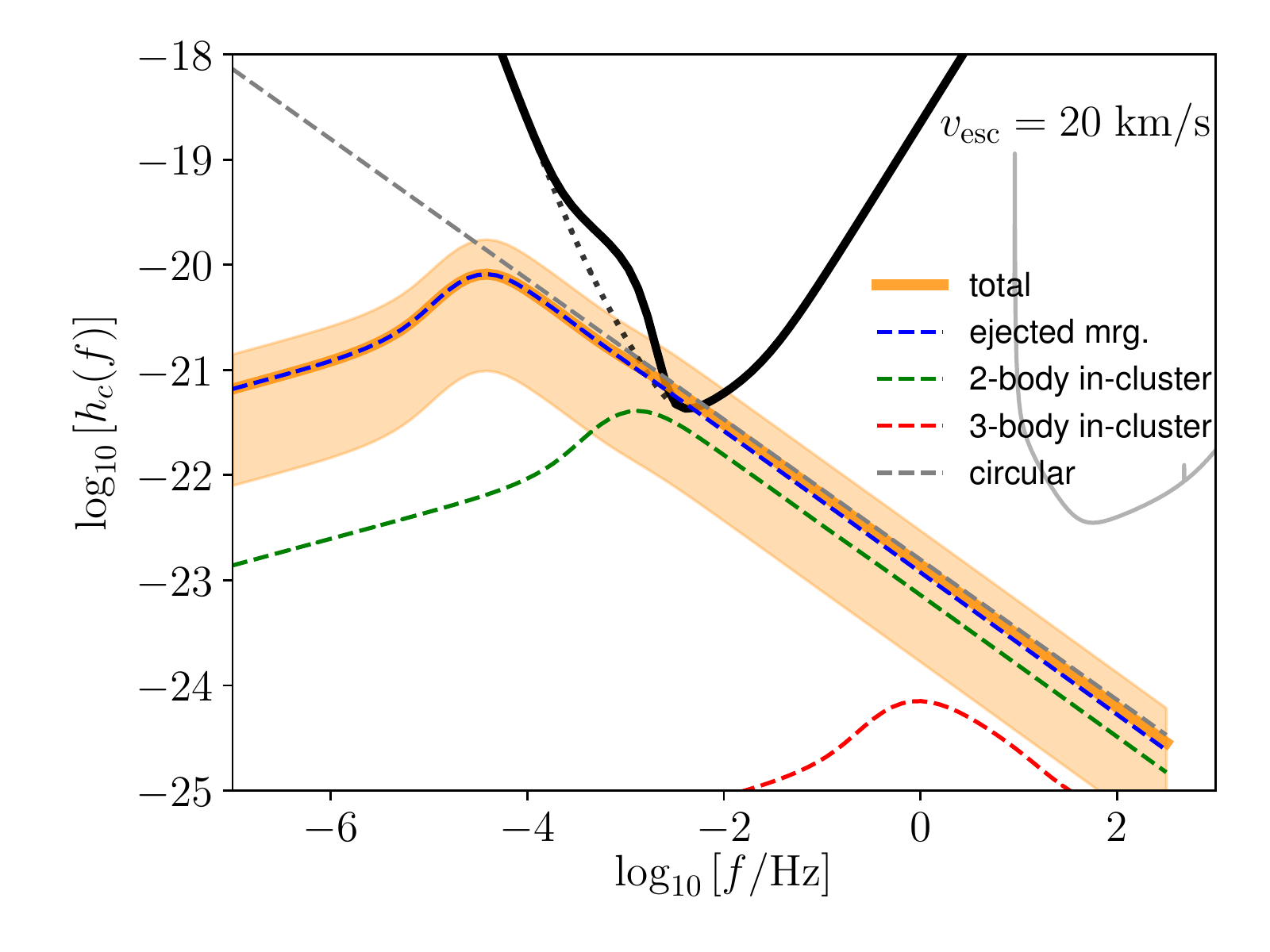} &
\end{array}$
\end{center}
\vspace{-15pt}
\caption{
The gravitational wave background of cluster-formed BBHs that will merge in
the LIGO band (\S \ref{S:GWB}). The left panel assumes GCs with an escape
velocity of $50$~km s$^{-1}$. The right panel assumes an escape velocity of
$20$~km s$^{-1}$. A merger rate of $100$~Gpc$^{-3}$yr$^{-1}$ is assumed with
an exception for the orange shaded region, which represents the total GWB for
the LIGO merger rate range $12-213$~Gpc$^{-3}$yr$^{-1}$.
}
% %
\label{Fig:GWB}
\end{figure*}

The gravitational wave background for a population of eccentric binaries has
characteristic amplitude \citep{Phinney:2001, Enoki+2007, Huerta+2015},
 \begin{equation}
 h^2_c(f) = \frac{G}{c^2} \frac{4}{\pi f} \int^{\infty}_0 \int^{\infty}_0 \int^{1}_{0}{ \frac{d^3n}{dz d\Mc de_0} \frac{dE}{df_r} \ de_0 d\Mc dz}.
 \label{Eq:GWB1}
\end{equation}
The first term under the integrand is the co-moving differential number
density of BBHs per redshift $z$, initial eccentricity $e_0$,
and chirp mass $\Mc$. The second term under the integrand is the energy emitted per unit
rest-frame-GW frequency, 
\begin{equation}
 \frac{dE}{df_r} = \sum^{n_{\max}}_{n=1}  \frac{dE_n}{df_r},
\end{equation}
where $dE_n/df_r$ is given by Eq. (\ref{Eq:dEndf}). 
At a fixed GW frequency, summing over $n = f_r/f_{\rm{orb}}$ is equivalent to
summing over every orbital frequency, throughout the lifetime of the binary,
that contributes power at the $n^{\rm{th}}$ harmonic to GWs at frequency
$f_r$. This can be understood graphically from Figure \ref{Fig:hctracks}. At a
given value of $f$ (and hence $z$), the sum over $n$ is the vertical sum of
all of the values of $h^2_c$ lying along the solid lines. The result is
characteristic bumps in the GWB for an initially eccentric binary (see our
Figures \ref{Fig:hctracks} and \ref{Fig:GWB}, and, \textit{e.g.},
\citet{CSD:2017}).

While the summation over $n$ is usually carried out up to infinity, we point
out that this is only valid when there is no specified orbital frequency of
formation. One should sum only over harmonics up to the maximum allowed
harmonic at a given GW frequency that corresponds to the initial orbital
frequency $n_{\max} = f/f_0$. For the values of $f_0$ relevant in his study,
we find that in the case of the 2-body and 3-body BBHs, this correction is
negligible. For example, for the ejected
population, we find that summing to infinity results in a $\sim5\%$
overestimate of the GWB at peak. Once the binaries circularize, this approximation
no longer affects our results.

We note that the form of Eq. (\ref{Eq:dEndf}), which is summed to compute the
GWB, is only valid for purely GW-driven decay of the binary. This is because
at each value of $n$ and $f$, one derives a new value of $f_{\orb}$ that
contributes to $f$ at the $n^{\rm{th}}$ harmonic. This value of $f_{\orb}$
along with $(e_0, \fint)$ is used to derive the corresponding value of $e$ at
that orbital frequency via Eq. (\ref{Eq:eof}). Because Eq. (\ref{Eq:eof}) is
specifically for GW-driven evolution, the derived spectrum is valid only for
GW-driven evolution as well. 

Additionally, Eq. (\ref{Eq:dEndf}) is only valid when the evolution time of
the BBHs is shorter than the Hubble time. This assumption holds for the 2-body
and 3-body populations, but it does not generally hold for the ejected
population. Future work is planned to extend the GWB calculation to include
redshift dependence in the BBH evolution in order to more accurately model the
GWB from the ejected population. This will also allow us to extend the GWB
calculation to the possibly large number of BBHs that will not merge in a
Hubble time, and are missed here and in most calculations of the GWB.

In the context of this study, where the correction for replacing $n_{\max}$
with infinity is small, where we assume no redshift dependence in BBH
evolution, and where GW decay is the only relevant binary evolution process,
we use the convenient fitting formula of \cite{CSD:2017} (their Eqs. 15 and
16) to replace all but the co-moving differential number density in our Eq.
(\ref{Eq:GWB1}).

To write the co-moving differential number density, we assume that the
co-moving merger rate measured by LIGO, $\mathcal{R}$, is the constant rate
out to a redshift of $z_{\max}=2$, and numerically  derive a probability
distribution for the initial orbital eccentricities, $P_i(e_0)$, from the
distributions derived in Paper I,
\begin{eqnarray}
\frac{d^3n}{dz d\Mc de_0} &=& k_i \mathcal{R}\frac{dt_r}{dz} P_i(e_0) \delta(\mathcal{M} - \mathcal{M}_0), \\ \nonumber
\frac{dt_r}{dz} &=& \left[H_0 (1+z) \sqrt{\Omega_M (1+z)^3 + (1+z) \Omega_k + \Omega_L} \right]^{-1} , 
\label{Eq:numden}
\end{eqnarray}
where unless specified, we take the LIGO merger rate to be
$\mathcal{R}\approx 100 \rm{yr}^{-1} \rm{Gpc}^{-3}$, we continue to consider a
distribution of BBHs with a single chirp mass $\Mc_0$, and the subscript $i$
denotes the $i^{th}$ formation channel, each of which makes up a fraction
$k_i$ of the entire population. We use a cosmology with
$H_0=70$kms$^{-1}$Mpc$^{-1}$, $\Omega_M=0.3$, and $\Omega_L=0.7$.

Because $dE/df_r$ encodes the entire orbital history of a binary with initial
orbital frequency $\fint$ and eccentricity $e_0$, one needs only to integrate
over the distribution of initial conditions ($e_0$, $\fint(e_0)$). These are
generated in Paper I. We truncate any contribution to the GWB below the
formation frequency $\fint(e_0)$ (which would be taken into account by the
value of $n_{\max}$ if we were not using $n_{\max}\rightarrow\infty$).

Then our calculation of the GWB, for the $i^{\rm{th}}$ population of BBHs, simplifies to
\begin{eqnarray}
h^2_{c,i}(f) &=&   k_i \mathcal{R}  \left(\frac{\Mc_0}{\Mc_{fit}}\right)^{5/3} \nonumber \\
&\times& \int^{z_{\max}}_{0} \Theta\left[f - \frac{f_0}{1+z}\right] \left(\frac{1+z}{1+z_{fit}}\right)^{-1/3}  \frac{dt_r}{dz} dz  \nonumber \\ 
&\times&  \int^{1}_{0} P_i(e_0) h^2_{c}(f, e_0, \fint) \ de_0,
\label{Eq:GWB_merge}
\end{eqnarray}
where $\Theta$ is the Heaviside step function and $h_{c}(f, e_0, \fint)$ is
given be Eqs. (13)-(15) in \citet{CSD:2017}. Corresponding to this fitting
formula are the values of $z_{fit}=0.02$ and $\Mc_{fit} = 4.16\times10^8
\Msun$. The total GWB is found from $h^2_c = \sum_{i} k_i h^2_{c,i}$. As in
Paper I, we assume a constant chirp mass of $\mathcal{M}_0 = 60 (1/2)^{6/5}$
for a $60 \Msun$ binary with a mass ratio of unity.

%%%%%%%%%%%%%%%%%%%%%%%%%%%%%%%%%%%%%%%%%%%%%%%%
%%% FIGURE %%%
%%%%%%%%%%%%%%%%%%%%%%%%%%%%%%%%%%%%%%%%%%%%%%%%    

\subsection{GWB: Results}

The fiducial GWB and its components are displayed in Figure
\ref{Fig:GWB}.  The thick orange line in Figure \ref{Fig:GWB} shows the total
GWB, calculated for BBHs formed in GCs, assuming a merger rate of
$100$~Gpc$^{-3}$yr$^{-1}$. The shaded orange region shows the range of
possible GWBs for the range of LIGO-inferred BBH merger rates
$12-213$~Gpc$^{-3}$yr$^{-1}$ \citep{GW170104}. The components of the GWB
associated with each cluster population are drawn as dashed lines of their
corresponding color: ejected (blue), 2-body (green), and 3-body (red). A
comparison background of circular- orbit field binaries matched to the LIGO
rate is plotted as a dashed grey line, \textit{i.e.}, the dashed grey line
would be the GWB if the LIGO rate corresponds to a population of circular-
orbit BBHs and the orange line is the GWB if the LIGO rate corresponds only to
GC formed BBHs.

In this section we focus on the left panel of Figure \ref{Fig:GWB}, which is
drawn for the fiducial GC parameters used in Paper I,
$v_{\rm{esc}}=50$~km s$^{-1}$ and single BH density of $10^5$ pc$^{-3}$.
In this scenario, the ejected, 2-body, and
3-body populations make up $51.2\%$, $46.4\%$, and $2.4\%$ of the entire GC
population, respectively.

The background of BBHs on circular orbits follows the expected $f^{-2/3}$ fall
of with GW frequency \citep[\textit{e.g.},][]{Phinney:2001, KocsisSesana:2011}.
The main feature prevalent in the eccentric GWB background are the bumps at
which the GWB rises to an equivalent circular-orbit GWB, but at lower
frequencies drops below the circular-orbit GWB. The peak of these bumps occurs
near the peak GW frequency associated with the average initial eccentricity of
each BBH population ($f^{\rm{peak}}(e_0)$). We note that the components of the
GWB associated with the merging BBH populations look very similar to the GWB
expected from a population with a single eccentricity and formation frequency
\citep[see][]{CSD:2017}. This is because the initial distributions derived in
Paper I are very narrow for each BBH population. Physically, this is because
the initial binary frequency is closely tied to the binary semi-major axis at
which a binary-single interaction would eject the binary from the cluster
($f_{\orb,0}\sim 10^{-7}$ for $v_{\rm{esc}}=50$~km s$^{-1}$, see Paper I).
Then because the small spread of initial orbital frequencies results in a
$\mathcal{T}\sim6 \times 10^{13}$ yr lifetime for a comparable circular-orbit
binary, and because we consider only the BBHs that will merge in the LIGO
band, it follows that the binary eccentricity at formation must be high in
order for the BBH to merge in a Hubble time (ejected, blue population), higher
still for the binary to merge before another cluster interaction (2-body,
green population), and even higher still to be captured and merge during a 3-body
interaction (3-body, red population). The combination of the two effects
selects for a very narrow range of initial binary parameters in each
population. See also discussions in Section \ref{sec:Eccentricity evolution}
and \ref{S:GCdep}.

The GWB of cluster-formed BBHs is largely indiscernible from a population of
circular-orbit BBHs above $10^{-3}$~Hz. At lower frequencies, the GWB exhibits
two bumps due to the 2-body BBHs at $10^{-3}$~Hz, and the ejected BBHs at
$10^{-4.5}$~Hz. These bumps occur below the LISA sensitivity curve. We
explore the dependence of the location and amplitude of these bumps on GC
properties in the next section.

Finally, the orange shaded region in Figure \ref{Fig:GWB} shows that,
for merger rates in the LIGO-inferred range that are above our fiducial value
of $100$~Gpc$^{-3}$yr$^{-1}$, the GWB falls just above the minimum of the
LISA sensitivity curve (consistent with the calculation for circular-orbit
binaries by \cite{SesanaLISALIGO:2016} and \citet{Nishizawa+2016}). We note,
however, that we have considered only $30+30 \Msun$ BBHs. In addition, we have not
computed the $S/N$ for a given LISA mission, which we leave to future work
that will consider a more detailed calculation of the GWB.

\section{Dependence on Globular Cluster Properties}
\label{S:GCdep}

In this section we explore the dependence of our results on GC properties,
focusing first on the GWB, and then applying our findings to the individually
resolved sources.

As shown in \S \ref{S:GWB}, the main difference between a GWB composed of a
dynamically-assembled eccentric population, and a canonical circular-orbit
field population is the appearance of characteristic bumps, as seen in Figure
\ref{Fig:GWB}. The location of a bump is roughly set by the frequency
associated with the maximum possible pericentre distance of the considered
population, which depends on the dynamical environment. It would be
interesting to determine if measurements of the location of one or more of
these bumps can be used to constrain the underlying BBH merger formation
channel. From Figure \ref{Fig:GWB}, one sees that for the ejected BBH mergers
(blue) the bump is generally located at frequencies too low to be observable
by LISA; however, the bump for the 2-body mergers is near the region where
LISA is most sensitive. In the following, we derive how the location of this 2
-body-merger bump and its relative normalization depends on the underlying GC
parameters.

Generally, the location of the GWB bump in frequency space will be near the
minimum peak GW frequency of that population, that is the peak GW frequency at
BBH formation. The reason is simply that below that frequency, the signal will
not only be suppressed, but also not build up as each BBH spirals towards
higher frequencies. In the following we derive how this minimum frequency, and
thereby the bump location, scales for the 2-body merger population.

For the 2-body BBH mergers the minimum peak GW frequency will be the peak 
GW frequency emitted by a BBH with semi-major axis equal to $a_{\rm ej}$ and with
GW inspiral life time, $\mathcal{T}(a_{\rm ej})$ equal to the time between
binary-single encounters, $t_{\rm bs}(a_{\rm ej})$. Using that
$\mathcal{T}(a_{\rm ej}) \approx t_{\rm c}(a_{\rm ej})(1-e^{2})^{7/2}$ (see
Section \ref{sec:Eccentricity evolution} and \cite{Peters64}), where $t_{\rm
c}(a_{\rm ej})$ refers to the BBH's GW lifetime assuming zero eccentricity,
one finds that the corresponding BBH (maximum) pericenter distance is given by
$r_{\rm p}/a_{\rm ej} \approx (1/2)(t_{\rm bs}(a_{\rm ej})/t_{\rm c}(a_{\rm
ej}))^{7/2}$. Assuming the relation $f \approx \pi^{-1}\sqrt{2Gm/r_{\rm
p}^{3}}$ presented in Eq. (\ref{Eq:fpeak}), and using that $t_{\rm bs}$ is the
inverse of the binary-single encounter rate \citep[e.g][]{2018ApJ...853..140S,
2017arXiv171107452S}, one now finds that the minimum peak GW frequency of the
2-body BBH merger population, $f_{\rm 2b}^{(\rm min)}$, can be written as,
\begin{equation}
{f_{\rm 2b}^{(\rm min)}} \approx 10^{-3.5} \text{Hz}\  \left(\frac{m}{30M_{\odot}}\right)^{2/7} \left(\frac{v_{\rm esc}}{50\text{kms}^{-1}}\right)^{-12/7} \left(\frac{n_{\rm s}}{10^{5}\text{pc}^{-3}}\right)^{3/7}.
\label{eq:fmin}
\end{equation}
To estimate the true location of the bump one has to further include the
redshift distribution of the merging BBHs, their assembly eccentricity
distribution, and how they merge inside their cluster. In our case one finds
that the bump is just a bit above the corresponding $f_{\rm 2b}^{(\rm min)}$,
as most of the 2-body BBHs indeed merge when their semi-major axis is $\approx
a_{\rm ej}$ (this is where $t_{\rm bs} \propto 1/a_{\rm ej}$ is at its maximum
and $\mathcal{T} \propto a_{\rm ej}^{4}$ is at its minimum), and most of the
sources that give rise to the highest signal are relative nearby.

As a result, from the relation given by Eq. (\ref{eq:fmin}) one concludes that
the 2-body population bump will move to the right when $v_{\rm esc}$ decreases
and $m,n_{\rm s}$ increase. As the mass term $m$ is not expected to change
more than of order unity and the corresponding bump location only scales
$\propto m^{2/7}$, one finds that the location is effectively a measure of
$v_{\rm esc}$ and $n_{\rm s}$, that is the underlying cluster properties.

The relative number of ejected mergers and 2-body mergers will also change
with $m,v_{\rm esc}$ and $n_{\rm s}$. To see how, we now derive an approximate
relation between the expected number of 2-body mergers and ejected mergers,
$N_{\rm 2b}/N_{\rm ej}$. For this, we first use that the probability for a BBH
to undergo a 2-body merger inside the cluster is $P_{\rm 2b} \approx P_{\rm
2b}(a_{\rm ej})/(1-\delta)$, where $P_{\rm 2b}(a_{\rm ej})$ is the probability
that a BBH with semi-major axis $a_{\rm ej}$ will merge before the next
binary-single encounter, and $\delta \approx 7/9$ is the average fractional
decrease in semi-major axis as a result of a hardening binary-single
interaction, as further explained in \cite{2017arXiv171107452S}. The number of
ejected mergers relate to the probability for an ejected BBH to merge within a
Hubble time, $P_{\rm H} \approx (t_{\rm H}/t_{\rm c}(a_{\rm ej}))^{2/7}$. As
$N_{\rm 2b}/N_{\rm ej} \approx P_{\rm 2b}/P_{\rm H}$, one now finds,
\begin{equation}
\frac{N_{\rm 2b}}{N_{\rm ej}} \approx 0.8 \times \left(\frac{m}{30M_{\odot}}\right)^{-4/7} \left(\frac{v_{\rm esc}}{50\text{kms}^{-1}}\right)^{6/7} \left(\frac{n_{\rm s}}{10^{5}\text{pc}^{-3}}\right)^{-2/7}.
\label{eq:N2}
\end{equation}
This relation shows how the normalizations of the GWB from the 2-body
population (characterized by a bump in the LISA band) and the ejected BBH
mergers (indistinguishable from simple power-law in the LISA band) scale
relative to each other.

Comparing the relation describing the bump location from Eq. (\ref{eq:fmin}),
and the expression for the relative number of 2-body mergers given by Eq.
(\ref{eq:N2}), one concludes that any variation in $m,v_{\rm esc}, n_{\rm s}$
that would move the bump towards higher frequencies into the LISA sensitivity
region, also will decrease the relative number of 2-body mergers. This
behavior can clearly be seen by comparing the panels in Figure \ref{Fig:GWB},
which present the GWB for a GC with $v_{\esc}=50$~km s$^{-1}$ (left) and
$v_{\esc}=20$~km s$^{-1}$ (right). The smaller escape velocity example causes
the GWB bump from the 2-body population to fall directly in the most sensitive
frequency range for LISA, but the smaller escape velocity also diminishes the
impact of the 2-body population on the GWB.

This implies, at least in our model, that it will become harder to observe the
bump the more it moves towards the frequency range to which LISA is most
sensitive. This complicates using the spectral shape of the GWB to constrain
the underlying BBH merger progenitor channel. However, it is worth
investigating more accurate  dynamical models to determine if this results
holds.

The above explanation for the GC-dependent location of the bumps
in the GWB applies also to the tracks of individually resolvable BBHs. Figure
\ref{Fig:hctracks2} shows the evolution of the three BBH populations at the
peak GW frequency for two different GC escape velocities, $v_{\esc}=50$~km
s$^{-1}$ (solid line) and $v_{\esc}=20$~km s$^{-1}$ (dashed line). As expected
from Eq. (\ref{eq:fmin}), the lower GC escape velocity causes the 2-body BBH
populations to emit at higher peak GW frequency than for the BBHs from higher
escape velocity GCs. The reason is that the BBHs in a lower escape velocity GC
will be ejected at a larger semi-major axis and hence must attain a larger
eccentricity in order to merge within the cluster. Higher eccentricity implies
a higher peak GW frequency. A similar argument holds for the ejected
population. The result, as seen in Figure \ref{Fig:hctracks2}, is that the
ejected and 2-body BBHs from the lower escape velocity GC have higher initial
peak GW frequencies, and hence circularize later.

The 3-body BBH mergers in Figure \ref{Fig:hctracks2} and \ref{Fig:GWB},
however, shows the opposite behavior; lower escape velocity GCs produce, on
average, relatively more eccentric 3-body mergers centered at lower
frequencies. The reason for the relative rate increase relates to how the
number of mergers contributing to each population scales with, in this case,
the escape velocity $v_{\esc}$. Following \cite{2017arXiv171107452S}, the
number of ejected mergers $\propto v_{\esc}^{16/7}$, the number of 2-body
mergers $\propto v_{\esc}^{22/7}$, where the number of 3-body mergers $\propto
v_{\esc}^{10/7}$. All three populations therefore decreases in number with
decreasing $v_{\esc}$, but the 3-body mergers decrease more slowly, which
leads to the observed relative increase compared to the two other populations.
Regarding the location at lower frequencies, this relates to the fact that the
characteristic distance for two BHs to undergo a GW capture in a resonating
three-body state (3-body merger) increases with the binary semi-major axis
$\propto a^{2/7}$, as derived in \cite{2014ApJ...784...71S,
2017arXiv171107452S}. A lower $v_{\esc}$ therefore leads to 3-body mergers to
generally form at larger pericenter distances, which also moves the population
to lower GW peak frequencies at formation.

\section{Conclusions}

We have examined the GW signatures of BBHs formed dynamically in GCs. In Paper
I of this series we presented three different formation channels for
dynamically formed BBHs that lead to three different BBH populations
identifiable by their eccentricities. Here we have computed the GW amplitude
and eccentricity evolution of representative BBHs from each population, as
well as a stochastic GW background from the entire population.

In order of increasing eccentricity at formation, the three merger populations
are `ejected mergers', `2-body mergers', and `3-body mergers', where the
ejected and the 2-body mergers each contribute equally to the observable rate,
while the 3-body mergers contribute up to $\sim5\%$. The ejected mergers
originate, as their name suggests, from BBHs that after $\mathcal{O}(20)$
binary-single scatterings have been ejected from their cluster through a
dynamical recoil. Those ejected with $e\gsim0.85$ will merge in a Hubble time,
and therefore contribute to the observable rate. The 2-body mergers, in
contrast, merge inside the cluster, in-between hardening binary-single
interactions before ejection can take place. These mergers will generally have
$e\gsim0.99$ at formation. The 3-body mergers form with even higher
eccentricities near $e\gsim0.9999$, as they are assembled during the
hardening binary-single interactions.

The ejected BBHs emit GWs at $\sim10^{-4.5}$~Hz at formation, just below
frequencies at which LISA is sensitive. The 2-body BBHs form with GW frequency
$\sim10^{-3}$~Hz, directly in the LISA band, and the 3-body BBHs form at
$\sim1$~Hz, in-between the LISA and LIGO bands.

For a 5-year LISA mission, we quantify the properties of the ejected and
2-body BBHs that will appear in the LISA band and merge in the LIGO band.
$30+30\Msun$ BBHs at a distance of 500 Mpc, that will merge in $\sim5$~yrs,
are detectable with a signal-to-nose of $\approx 10$ in LISA starting at a GW
frequency of $\sim10^{-1.8}$~Hz. At this frequency the average ejected BBH has
an eccentricity of $e\sim8 \times 10^{-4}$ while the average 2-body BBH has an
eccentricity of $e\sim 10^{-2}$. Because LISA can measure eccentricities above
$\sim10^{-3}$ \citep{Nishizawa+2016}, and because the eccentricity
distribution of both populations does not significantly overlap at this GW
frequency (Paper I), this means that the 2-body and ejected populations will
be discernible by LISA, via their eccentricities
\citep[see][]{Nishizawa+2017}. Furthermore, LISA will be able to watch the
eccentricity of the 2-body BBHs evolve by an order of magnitude as they leave
the LISA band, at $\sim0.1$~Hz. In contrast, the ejected population has
eccentricities marginally detectable by LISA when leaving the band.

The 3-body population cannot be detected by LISA, which is a clue to BBH
formation on its own \citep{SamsingDOrazio:2018a}. However, the 3-body BBHs
will enter the LIGO band with a significant eccentricity of $e\sim0.1$. We
have shown here that despite their eccentricity, these BBH systems should be
delectable in advanced LIGO, and also could have been detected by LIGO O1.
That no eccentric BBHs were detected in LIGO O1 and O2 is consistent with this
population making up only $\sim5\%$ of dynamically formed BBHs
\citep{2017arXiv171107452S}. Future LIGO observations will begin to constrain
the fraction of BBHs that are formed dynamically. A detection of highly
eccentric BBHs making up $\lesssim 5\%$ of BBH mergers in LIGO would lend
significant evidence towards the BBH formation picture presented in Paper I
and in other works \citep[\textit{e.g.},][]{2017arXiv171107452S, RodPND+2017}.

We compute the gravitational wave background (GWB) generated by the
eccentric GC-formed BBHs. We find characteristic bumps in the eccentric GWB,
the GW frequency of which depends most strongly on the GC escape velocity.
While low escape velocity clusters cause such a GWB bump from the 2-body
population to fall directly in the LISA band, the same low escape velocity
clusters also cause the GWB due to ejected mergers to dominate over the 2-body
GWB. Hence the GWB is very similar to that found for BBHs on circular orbits
formed in the field \citep[\textit{e.g.},][]{SesanaLISALIGO:2016,
Nishizawa+2016}. In any case, the GWB is marginally detectable for the
fiducial BBH populations considered here.

As we have shown, the GW signatures of eccentric BBHs formed in GCs provide a
powerful tool for deciphering the origin of BBH systems, a tool uniquely
suited for multi-band GW astronomy of the coming decade. In future work we
plan to expand upon the schematic calculation of the GWB presented here.
Specifically, we wish to extend our calculation to more accurately capture the
GWB from the ejected populations, and to the BBHs that stall, with a range of
eccentricities below that needed to merge. To achieve this one must adapt the
commonly used expression for the GWB strain (Eq. \ref{Eq:GWB1}), to include
BBH evolution on a timescale that is comparable to the Hubble time.
Additionally, a more accurate representation of the distribution of chirp
masses and the redshift distribution must be included, not only for the GWB
calculation but most importantly to estimate the number and orbital parameter
distribution of BBHs that LISA and LIGO will detect.

\vspace{-10pt}

\section*{Acknowledgements}
We thank the referee, Emanuele Berti, for constructive suggestions that
enhanced the quality of this work. We thank Alberto Sesana and Bence Kocsis
for helpful discussions pertaining to LISA and the signal-to-noise
estimates.
%We wish we could publicly thank the great american turkey vulture for inspiration.
Financial support was provided from NASA through Einstein Postdoctoral
Fellowship award number PF6-170151 (D.J.D.) and from the
Lyman Spitzer Fellowship (J.S.). 

%%%%%%%%%%%%%%%%%%%%%%%%%%%%%%%%%%%%%%%%%%%%%%%%%%

%%%%%%%%%%%%%%%%%%%% REFERENCES %%%%%%%%%%%%%%%%%%
\bibliography{LISABBH_refs}
\bibliographystyle{mnras}
%%%%%%%%%%%%%%%%%%%%%%%%%%%%%%%%%%%%%%%%%%%%%%%%%%

% Don't change these lines
\bsp	% typesetting comment
\label{lastpage}
\end{document}